\documentclass[
 prl,
 reprint,
 floatfix,
 amsmath,amssymb,
 showpacs,
 superscriptaddress,
 citeautoscript,
 nonatbib
]{revtex4-2}

\usepackage{xr}
\makeatletter
\newcommand*{\addFileDependency}[1]{
  \typeout{(#1)}
  \@addtofilelist{#1}
  \IfFileExists{#1}{}{\typeout{No file #1.}}
}
\makeatother

\usepackage{amssymb}

\usepackage[breaklinks]{hyperref}
\usepackage[all]{hypcap}
\hypersetup{
    plainpages=false,
    unicode=false,                          
    pdfborder={0 0 0},
    pdftoolbar=true,                        
    pdfmenubar=true,                        
    pdffitwindow=false,                     
    pdfstartview={FitH},                    
    pdftitle={},                            
    pdfauthor={},                           
    pdfproducer={LNCMI-CNRS-UGA-UPS-INSA},  
    pdfkeywords={High} {Magnetic} {Fields}, 
    pdfnewwindow=true,                      
    linktoc=section,
    colorlinks=true,                        
    linkcolor=red,                         
    citecolor=red,                          
    filecolor=magenta,                      
    urlcolor=blue                           
}

\usepackage{graphicx}
\usepackage{float}
\usepackage{amsmath}
\usepackage{bm}
\usepackage{color, soul}
\usepackage[normalem]{ulem}
\usepackage{siunitx}
\usepackage{upgreek}
\usepackage[usenames,dvipsnames]{xcolor}
\usepackage{textcomp}
\usepackage{chemformula}
\usepackage[normalem]{ulem}

\usepackage{booktabs}
\usepackage{amssymb}
\usepackage{colortbl}

\usepackage{pgfplots}
  \pgfplotsset{compat=newest}
  
  \usetikzlibrary{plotmarks}
  \usetikzlibrary{arrows.meta}
  \usepgfplotslibrary{patchplots}

\makeatletter
\AtBeginDocument{\let\hl\@firstofone}
\makeatother

\begin{document}

\title{Correlated Anharmonicity and Dynamic Disorder \\Control Carrier Transport in Halide Perovskites}

\author{Maximilian J. Schilcher}
\affiliation{Physics Department, TUM School of Natural Sciences, Technical University of Munich, 85748 Garching, Germany}
\author{David J. Abramovitch}
\altaffiliation{present address: Department of Applied Physics and Materials Science, California Institute of Technology, Pasadena, California 91125, United States}
\affiliation{Department of Physics, University of California Berkeley, Berkeley, California 94720, United States}
\author{Matthew Z. Mayers}
\altaffiliation{present address: Google, Los Angeles, California 90291, United States}
\affiliation{Department of Chemistry, Columbia University, New York, New York 10027, United States}
\author{Liang Z. Tan}
\affiliation{The Molecular Foundry, Lawrence Berkeley National Laboratory, Berkeley, California 94720, United States}
\author{David R. Reichman}
\affiliation{Department of Chemistry, Columbia University, New York, New York 10027, United States}
\author{David A. Egger}
\email{david.egger@tum.de}
\affiliation{Physics Department, TUM School of Natural Sciences, Technical University of Munich, 85748 Garching, Germany}

\begin{abstract}
\noindent Halide pervoskites are an important class of semiconducting materials which hold great promise for optoelectronic applications.  In this work we investigate the relationship between vibrational anharmonicity and dynamic disorder in this class of solids.  Via a multi-scale model parameterized from first-principles calculations, we demonstrate that the non-Gaussian lattice motion in halide perovskites is microscopically connected to the dynamic disorder of overlap fluctuations among electronic states.  This connection allows us to rationalize the emergent differences in temperature-dependent mobilities of prototypical MAPbI$_3$ and MAPbBr$_3$ compounds across structural phase-transitions, in agreement with experimental findings.  Our analysis suggests that the details of vibrational anharmonicity and dynamic disorder can complement known predictors of electronic conductivity and can provide structure-property guidelines for the tuning of carrier transport characteristics in anharmonic semiconductors.
\end{abstract}

\date{\today}

\maketitle

\begin{figure}
    \centering
    \includegraphics[scale = 0.42, trim=0mm 0mm 0mm 0mm, clip]{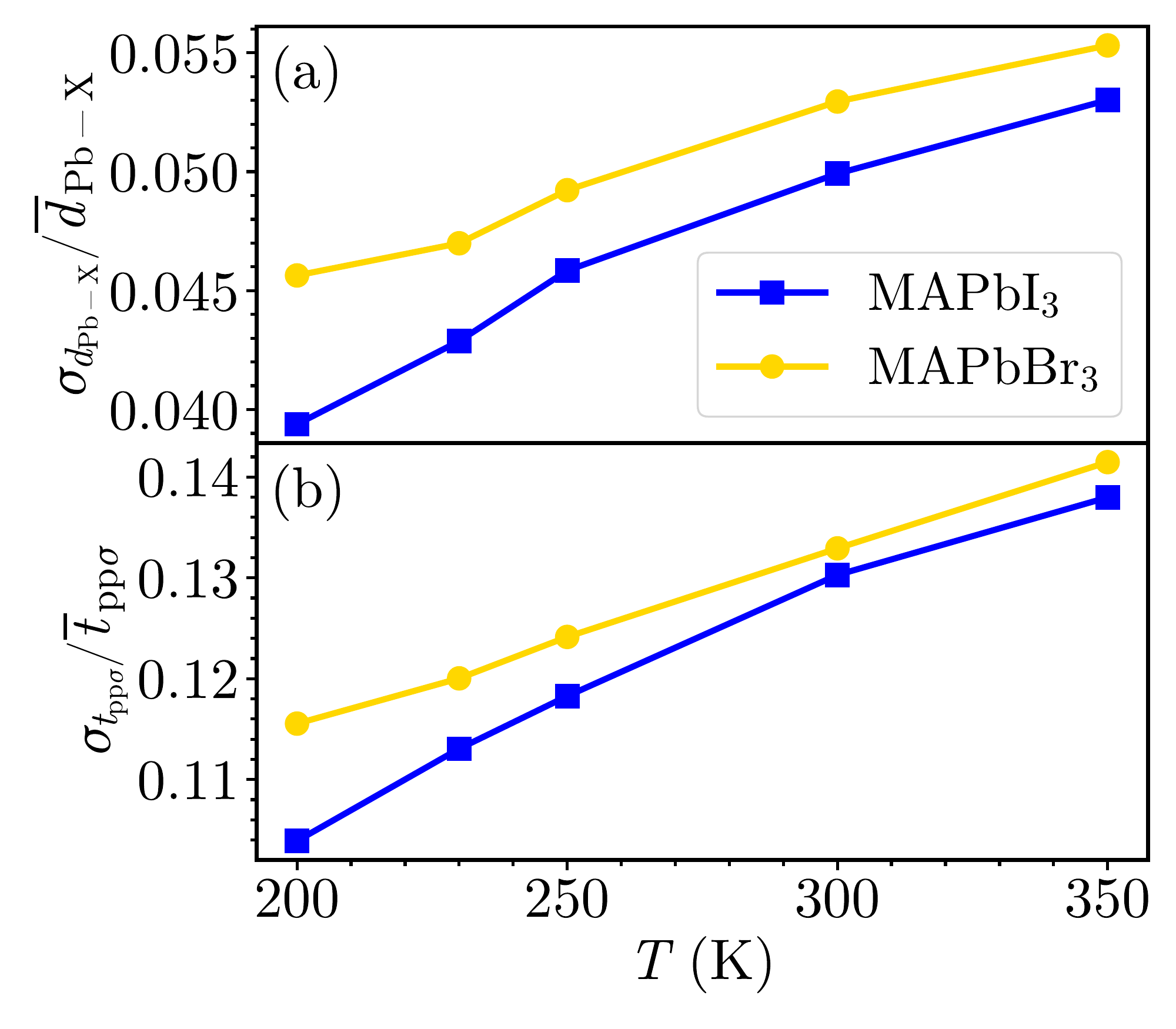}
    \caption{Relative fluctuations in Pb-\textit{X} bond-distances and $t_\mathrm{pp\sigma}$ overlaps for both materials as a function of temperature.
    Data calculated as the ratio between the standard deviation ($\sigma_{\mathrm{d}_\mathrm{Pb-X}}$ and $\sigma_\mathrm{pp\sigma}$, respectively) and mean value ($\overline{d}_\mathrm{Pb-X}$ and $\overline{t}_\mathrm{pp\sigma}$, respectively).}
    \label{fig:4}
\end{figure}

\noindent Halide perovskites (HaPs) are crystalline semiconductors that are relevant for a variety of technological applications, in particular as photovoltaic materials \cite{snaith_PerovskitesEmergenceNew_2013, green_EmergencePerovskiteSolar_2014, stranks_MetalhalidePerovskitesPhotovoltaic_2015, correa-baena_PromisesChallengesPerovskite_2017, li_ChemicallyDiverseMultifunctional_2017, nayak_PhotovoltaicSolarCell_2019}. 
The favorable device characteristics of HaPs are seemingly rooted in their optoelectronic properties  \cite{brenner_HybridOrganicInorganic_2016, stoumpos_HalidePerovskitesPoor_2016a}.
In particular, they possess direct band gaps and exciton binding energies smaller than thermal energy at ambient conditions. These factors enable strong sunlight absorption and rapid separation of electrons and holes in HaP thin films.  Furthermore, the low carrier effective-masses in these materials signal efficient electronic transport.  Together with low non-radiative recombination rates \cite{stranks_ElectronHoleDiffusionLengths_2013}, these properties enable efficient capture of light-generated carriers at the contacts.\\ 
Interest in HaPs as a promising material platform is heightened by their tunability.  In particular, chemical variation across the \textit{A}, \textit{B}, and \textit{X} ions of their $ABX_3$ stoichiometry can, in principle, create a knob with which to alter their properties with seemingly small changes in their overall structure \cite{saparov_OrganicInorganicPerovskites_2016}. 
Indeed, the electronic, vibrational and dielectric properties of HaPs can be adjusted \textit{via} tailoring their ionic composition even in high-symmetry HaP phases \cite{brenner_HybridOrganicInorganic_2016, stoumpos_HalidePerovskitesPoor_2016a,saparov_OrganicInorganicPerovskites_2016}.
This is relevant technologically since it enables, \textit{e.g.}, control over the fundamental band gap which can be used to increase power-conversion efficiencies of HaP tandem solar cells \cite{lal_PerovskiteTandemSolar_2017}.\\ 
However, the predictive power of established structure-property relationships is challenged in HaPs because their finite-temperature properties are unusual among optoelectronic materials \cite{egger_WhatRemainsUnexplained_2018a}, especially with respect to their charge transport characteristics. 
Experiment and theory agree that carrier mobilities around room temperature are limited by phonon scattering~\cite{herz_ChargeCarrierMobilitiesMetal_2017a}. 
However several contradictions between experimental data and predictions from standard transport theories remain unexplained \cite{schilcher_SignificancePolaronsDynamic_2021}. 
Indeed, the confluence of large amplitude, anharmonic atomic displacements \cite{beecher_DirectObservationDynamic_2016, yaffe_LocalPolarFluctuations_2017a, wu_LightinducedPicosecondRotational_2017, yang_SpontaneousOctahedralTilting_2017, gold-parker_AcousticPhononLifetimes_2018a, klarbring_LowenergyPathsOctahedral_2019, liu_ThermalDisorderBond_2019, ferreira_DirectEvidenceWeakly_2020, lanigan-atkins_TwodimensionalOverdampedFluctuations_2021, debnath_CoherentVibrationalDynamics_2021, cannelli_AtomicLevelDescriptionThermal_2022, zhu_ProbingDisorderCubic_2022, fransson_LimitsPhononQuasiparticle_2023, weadock_NatureDynamicLocal_2023a}
in a polar lattice \textit{and} dispersive electronic band structures \cite{brenner_HybridOrganicInorganic_2016, stoumpos_HalidePerovskitesPoor_2016a, herz_ChargeCarrierMobilitiesMetal_2017a} 
introduces behavior that is difficult to capture in standard theoretical models \cite{schilcher_SignificancePolaronsDynamic_2021, irvine_QuantifyingPolaronicEffects_2021, zhang_EffectQuarticAnharmonicity_2022, zacharias_AnharmonicElectronphononCoupling_2023, seidl_AnharmonicFluctuationsGovern_2023}.\\
Specifically, HaPs \hl{have been discussed to }feature ultra-short carrier relaxation times and mean-free paths on the order of \hl{only a few unit cells as shown experimentally and theoretically} \cite{karakus_PhononElectronScattering_2015a, hill_PerovskiteCarrierTransport_2018, lacroix_ModelingElectronicMobilities_2020a}, which violates the Mott-Ioffe-Regel (MIR) criterion and renders the most widely-used versions of standard kinetic theory inapplicable \cite{ioffe_NoncrystallineAmorphousLiquid_1960, mott_ConductionNoncrystallineSystems_1972, schilcher_SignificancePolaronsDynamic_2021}.
\hl{Related to this, recent experimental}~\cite{biewald_TemperatureDependentAmbipolarCharge_2019a}\hl{ and theoretical studies}~\cite{mayers_HowLatticeCharge_2018,lacroix_ModelingElectronicMobilities_2020a,schilcher_SignificancePolaronsDynamic_2021}\hl{ have highlighted the shortcomings of Boltzmann transport approaches in explaining the charge transport characteristics of HaPs that have been established experimentally}~\cite{herz_ChargeCarrierMobilitiesMetal_2017a}.
Supporting this viewpoint, high-level numerical treatments confirm that in the Fr\"ohlich polaron model a quasiparticle-based momentum representation of charge carriers is inadequate in the intermediate coupling regime of relevance for semiconductors such as HaPs \cite{mishchenko_PolaronMobilityQuasiparticles_2019}\hl{: it was shown that for the intermediate coupling regime ($\alpha=2.5$) the MIR limit is violated in the Fr\"ohlich polaron model over a range of $0.2<k_\mathrm{B}T/\hbar\omega<10$, with $\hbar\omega$ being an optical phonon energy.}~\cite{mishchenko_PolaronMobilityQuasiparticles_2019}
\hl{Using $\hbar\omega_\mathrm{LO}{\approx}15$\,meV for MAPbI$_3$} \cite{wright_ElectronPhononCoupling_2016}\hl{, this translates into a wide temperature range of 30\,K to 1740\,K where the MIR limit is violated and standard kinetic theory does not apply, as more recently re-emphasized in Ref.~}\cite{martin_MultiplePhononModes_2023}.\\
\hl{In this context, it is interesting that} the lattice dynamics in HaPs are localized in real space because of strong anharmonicity \cite{yaffe_LocalPolarFluctuations_2017a, ferreira_DirectEvidenceWeakly_2020, weadock_NatureDynamicLocal_2023a}. 
\hl{This type of vibrational anharmonicity occurs when the atomic motions in the system enter regimes of the potential energy surface that deviate from the harmonic approximation.
However, traditional approaches to both the Fr\"ohlich polaron model and the Boltzmann transport equation employ the harmonic approximation.
Together with the aforementioned further shortcomings of traditional kinetic methods to describe carrier scattering in HaPs that have been discussed in the literature, this motivates us to explore a real-space theoretical approach that leaves aside a purely particle-like momentum-space representation of carriers.
Parametrizing such a method from first-principles and comparing the mechanism of charge transport across related materials enables us to detect how the transient localization of carriers influences their mobility.}\\
Previous work by several of the present authors on the prototypical variant MAPbI$_3$ demonstrated that for near room temperature conditions, \textit{dynamic disorder} is prevalent. Namely, large atomic displacements induce strong fluctuations in electronic overlaps, which dictate carrier mobility and its temperature-dependence \cite{mayers_HowLatticeCharge_2018}. 
Lacroix \textit{et al.} found that a Fr\"ohlich-type scattering, where strong disorder induces localization of charge, is consistent with measured carrier diffusion coefficients and experimentally-measured mobility magnitudes \cite{lacroix_ModelingElectronicMobilities_2020a}.
Both studies centered on mechanisms where the modulation of the electronic couplings by anharmonic atomic displacements, which have been found to be substantially nonlinear in MAPbI$_3$ \cite{mayers_HowLatticeCharge_2018, abramovitch_ThermalFluctuationsCarrier_2021}, are used to predict transport properties.
However, the precise connections between vibrational anharmonicity and dynamic disorder are not known, despite their relevance for various systems, including organic \cite{troisi_ChargeTransportRegimeCrystalline_2006, troisi_ChargeTransportHigh_2011a, wang_ChargeTransportOrganic_2013, fratini_TransientLocalizationScenario_2016, fratini_MapHighmobilityMolecular_2017, asher_AnharmonicLatticeVibrations_2020a, giannini_FlickeringPolaronsExtending_2020, stoeckel_AnalysisExternalInternal_2021,  giannini_TransientlyDelocalizedStates_2023} 
and ionic semiconductors, \textit{e.g.}, SrTiO$_3$ \cite{zhou_PredictingChargeTransport_2019}. 
Since carrier scattering by phonons is a limiting mechanism for electronic transport close to room temperature, rationalizing the underlying microscopic origins and connections between anharmonicity and dynamic disorder is clearly required for the development of predicitive structure-property relationships for HaPs and similarly for a broader class of anharmonic semiconductors.
\hl{One way to establish such connections is \textit{via} comparison of related but distinct material compounds in regard to their dominant scattering mechanisms and charge-transport behavior.}\\
In this letter, we investigate carrier dynamics in the prototypical anharmonic HaP semiconductors MAPbI$_3$ and MAPbBr$_3$ through a multi-scale theoretical model that is parameterized from first-principles calculations.
Analyzing the temperature-dependent vibrational anharmonicity, it is found that MAPbBr$_3$ is significantly \textit{more anharmonic} at lower temperatures, in line with \hl{what can be expected} from its lower tetragonal-to-cubic phase-transition temperature.
We show that MAPbBr$_3$ has a reduced carrier mobility compared to MAPbI$_3$ in this temperature range because of the stronger anharmonicity of its lattice, which results in a weaker mobility temperature dependence overall.
A spectral analysis of the dynamic disorder provides precise connections to anharmonicity, since both effects become more similar in the two compounds as temperature increases, until carrier mobilities are comparable.
Our work supports a transient localization-type picture of carrier mobility in HaPs, where carrier diffusion follows atomic vibrations.  It is demonstrated that carrier mobilities can be altered through anharmonicity and dynamic disorder, establishing these effects as handles for tuning transport properties in an important class of semiconductors.\\\begin{figure}[t]
    \centering
    \includegraphics[scale = 0.42, trim=0mm 0mm 0mm 0mm, clip]{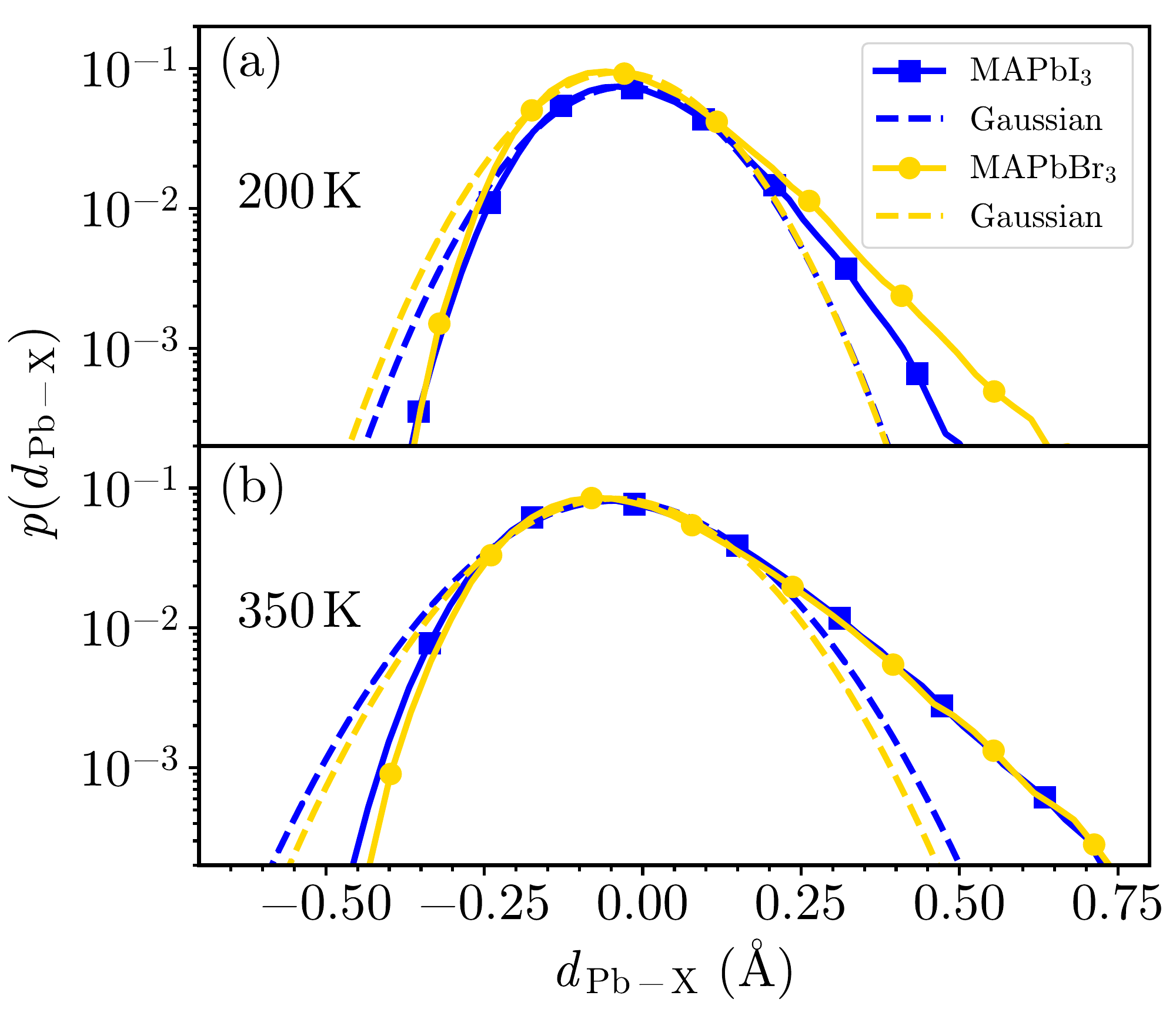}
    \caption{Histograms of the Pb-\textit{X} bond-distances in MAPbI$_3$ and MAPbBr$_3$ at 200\,K (panel a) and 350\,K (b), computed \textit{via} force-field-based MD calculations.
    The dashed lines are Gaussian fits to the respective distributions, where deviations to the actual data signify vibrational anharmonic effects. 
    \hl{The mean value of Pb-\textit{X} bond-distances is set to zero in all plots. Histograms have been normalized by dividing each data point by the total number of points (number of bins: 50).}}
    \label{fig:1}
\end{figure}%
We perform molecular dynamics (MD) calculations of MAPbI$_3$ and MAPbBr$_3$ to account for anharmonic vibrations at various temperatures that include the tetragonal and cubic phase of both materials.
Specifically, we apply previously-reported force fields \cite{mattoni_MethylammoniumRotationalDynamics_2015, hata_DevelopmentClassicalInteratomic_2017} 
in order to enable large-scale/long-time MD calculations of $16\times16\times16$ supercells (49152 atoms) with \texttt{LAMMPS} \cite{plimpton_FastParallelAlgorithms_1995} (see Ref.~\cite{SM} for details).
\hl{Notably, the force-field MD calculations include anharmonic effects because they were shown to capture phenomena in MAPbI$_3$ and MAPbBr$_3$ that are explicitly anharmonic, \textit{e.g.}, temperature-induced lattice expansions and phase-transitions}~\cite{mattoni_MethylammoniumRotationalDynamics_2015,hata_DevelopmentClassicalInteratomic_2017}.\\
Fig.~\ref{fig:1} shows histograms of computed Pb-\textit{X} bond-distances of the two compounds at 200\,K and 350\,K that were extracted from $NVT$-MD production runs following extensive $NpT$-MD equilibration.
At 200\,K, the Pb-Br bond-distance distribution is significantly more non-Gaussian than its Pb-I counterpart. In particular, the histograms reveal that deviations from Gaussian behavior for the larger-distance displacements in MAPbBr$_3$ are significantly more prominent at that temperature.
\hl{This can be quantified by calculating the ratios of the standard deviations of the recorded Pb-\textit{X} bond-distance distribution and the Gaussian fit. At 200\,K, they are found to be 1.04 and 1.12 for MAPbI$_3$ and MAPbBr$_3$, respectively, confirming that the latter is deviating more from the harmonic behavior.
The finding agrees} with expectations borne from the substantially lower tetragonal-to-cubic phase-transition temperature in MAPbBr$_3$ (${\approx}240$\,K) compared to MAPbI$_3$ (${\approx}330$\,K) \cite{poglitsch_DynamicDisorderMethylammoniumtrihalogenoplumbates_1987}. 
\hl{It can be rationalized by the larger ionic radius of iodine, which implies that a higher thermal energy is required for reaching an on-average cubic symmetry of MAPbI$_3$ compared to MAPbBr$_3$.}\\
\hl{In line with this expectation and our findings}, previous work found that MAPbI$_3$ features a potential surface that is significantly more anharmonic in the cubic than in the tetragonal phase, where large-amplitude anharmonic displacements accompanying octahedral tiltings are confined to occur only in two spatial dimensions \cite{sharma_ElucidatingAtomisticOrigin_2020}. 
Furthermore, recent neutron scattering experiments comparing the two compounds found that the disorder correlation-length is significantly shorter in MAPbBr$_3$ at lower temperature, in line with our findings \cite{weadock_NatureDynamicLocal_2023a}. 
Accordingly, above the phase-transition temperature of MAPbI$_3$ at {340\,K}, when both materials are in the cubic phase, differences in the bond-distance distributions are minor and the two compounds are similarly anharmonic (see Fig.~\ref{fig:1}).
\hl{Calculating the ratios of the standard deviations of the recorded Pb-\textit{X} bond-distance distribution and the Gaussian fit like above, we find them to be 1.08 and 1.12 for MAPbI$_3$ and MAPbBr$_3$, respectively, confirming that at 350\,K the degree of anharmonicity in both compounds is more similar than at 200\,K.}
The differences in anharmonic vibrational behaviors of MAPbI$_3$ and MAPbBr$_3$ at lower and higher temperatures allows for a determination of the impact of this effect on finite-temperature electronic structure and carrier dynamics.\\
We determine the finite-temperature electronic properties through a multi-scale tight-binding (TB) model (see \cite{SM} and Ref.~\cite{mayers_HowLatticeCharge_2018} for details) that is parameterized \textit{via} first-principles MD and one-shot Wannier projections onto a local atomic basis \cite{mostofi_UpdatedVersionWannier90_2014}, both using density functional theory (DFT) as implemented in \texttt{VASP} \cite{kresse_EfficientIterativeSchemes_1996} and \texttt{Quantum Espresso} \cite{giannozzi_UantumESPRESSOExascale_2020}. 
Importantly, this TB model is sensitive to structural fluctuations \textit{via} inclusion of distance-dependent onsite and overlap terms in the Hamiltonian which are fitted using DFT-based MD.
The model employs temperature-dependent trajectories from force-field-based MD to obtain statistical information on the finite-temperature electronic structure and uses this information in conjunction with quantum-dynamical simulations of the carrier dynamics.
The latter are performed using an Ehrenfest approach that neglects the back reaction forces on the lattice , applied on  $96\times96\times96$ real-space supercell Pb-\textit{X} motifs.
The impact of back reaction forces on the carrier scattering is expected to be small: formation of a Fr\"ohlich polaron would require coherent long wavelength vibrations whereas in HaPs the relevant lattice dynamics are localized in real space.\\
\begin{figure}[t]
    \centering
    \includegraphics[scale = 0.42, trim=0mm 0mm 0mm 0mm, clip]{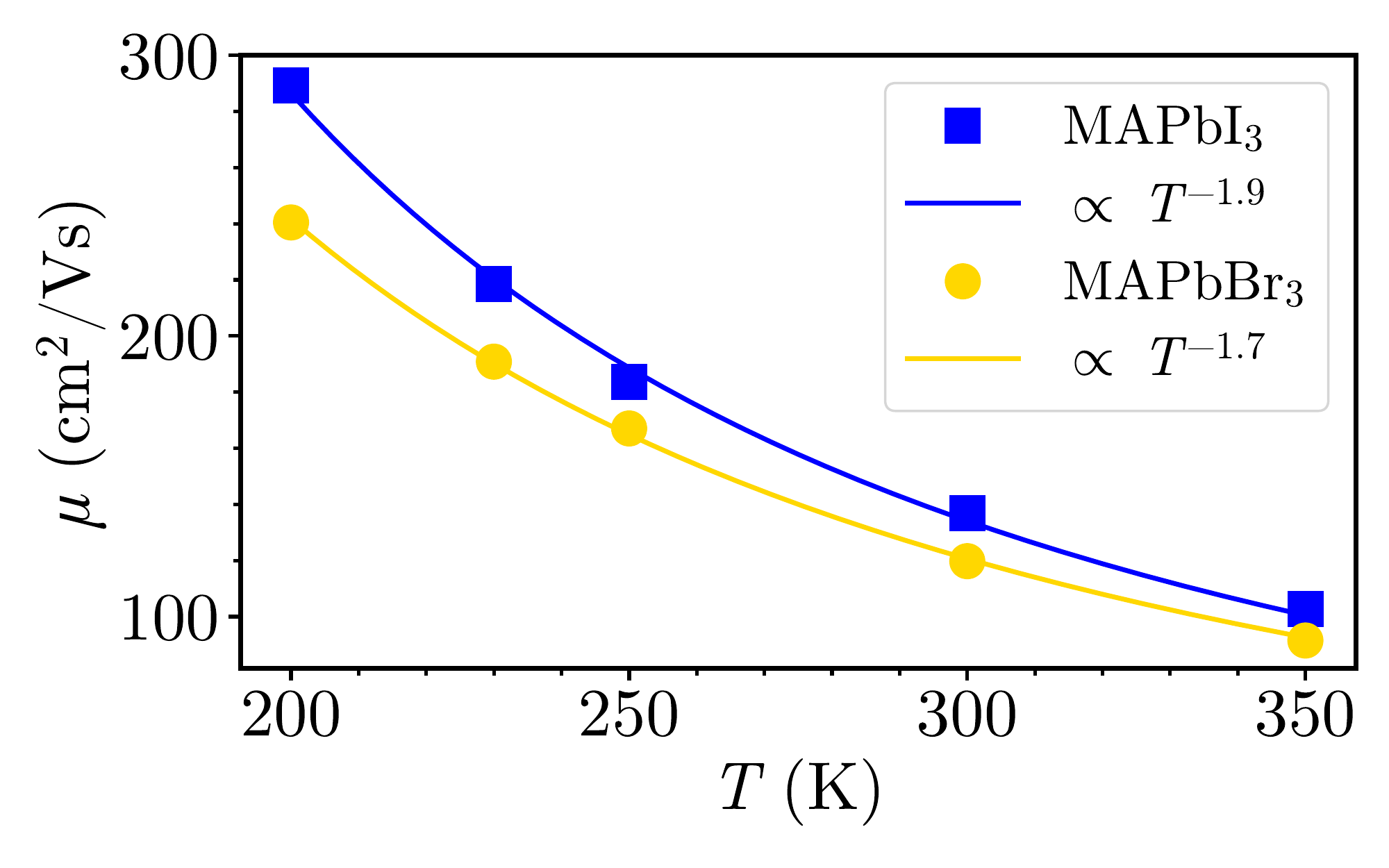}
    \caption{Temperature-dependent charge carrier mobilities (sum of electrons and holes) for MAPbI$_3$ and MAPbBr$_3$ computed \textit{via} our multi-scale TB model and quantum dynamics approach.
    The lines represent best fits to the power-law behavior of the temperature-dependent mobility data.}
    \label{fig:2}
\end{figure}%
The resulting temperature-dependent carrier mobilities of MAPbI$_3$ and MAPbBr$_3$ are shown in Fig.~\ref{fig:2}.
In the region where MAPbBr$_3$ was found to exhibit more profound anharmonicity than MAPbI$_3$ (200-300\,K, \textit{cf.} Fig.~\ref{fig:1}), its mobility is reduced and its temperature dependence is flatter.
When temperature is increased, progressively more anharmonic displacements appear in MAPbI$_3$ and the temperature-dependence of its mobility is concomitantly altered.
Close to room temperature, where MAPbI$_3$ is still in the tetragonal phase, its carrier mobility remains noticeably higher than the one of MAPbBr$_3$.
Interestingly, at 350\,K the carrier mobilities of the compounds are comparable, since both are in the cubic phase and their atomic dynamics are similarly anharmonic (\textit{cf.} Fig.~\ref{fig:1}).\\
The observed power-law behaviors of the mobilities (see Fig.~\ref{fig:2}) are in broad agreement with experimental observations \cite{herz_ChargeCarrierMobilitiesMetal_2017a,oga_ImprovedUnderstandingElectronic_2014, savenije_ThermallyActivatedExciton_2014, milot_TemperatureDependentCharge_2015, karakus_PhononElectronScattering_2015a, yi_IntrinsicChargeTransport_2016a, shrestha_AssessingTemperatureDependence_2018a, biewald_TemperatureDependentAmbipolarCharge_2019a, xia_LimitsElectricalMobility_2021, bruevich_IntrinsicTrapFree_2022}.
In particular, the room temperature mobility magnitudes and the finding that MAPbI$_3$ is more conductive than MAPbBr$_3$ at that temperature match well with recent experimental findings~\cite{chen_OptoelectronicPropertiesMixed_2022}.
It is noted that perfect agreement between theory and experiment, both for mobility magnitudes and temperature dependencies, cannot be expected because of experimental variations induced by sample fabrication and characterization methods \cite{herz_ChargeCarrierMobilitiesMetal_2017a} as well as neglect of certain mechanisms, \textit{e.g.}, defect scattering,  in our model.
\hl{Furthermore, our model applies approximate treatments to calculate electronic properties and their dependencies on structural fluctuations, which may lead to additional inaccuracies. 
The finding that our approach correctly captures the changes of the mobility characteristics when comparing MAPbI$_3$ and MAPbBr$_3$ signifies that the model accounts for the carrier scattering mechanisms that determine charge transport behavior in these materials.
In the following, we will provide a detailed description of these mechanisms.}\\ 
\begin{figure}[b]
    \centering
    \includegraphics[scale = 0.42, trim=0mm 0mm 0mm 0mm, clip]{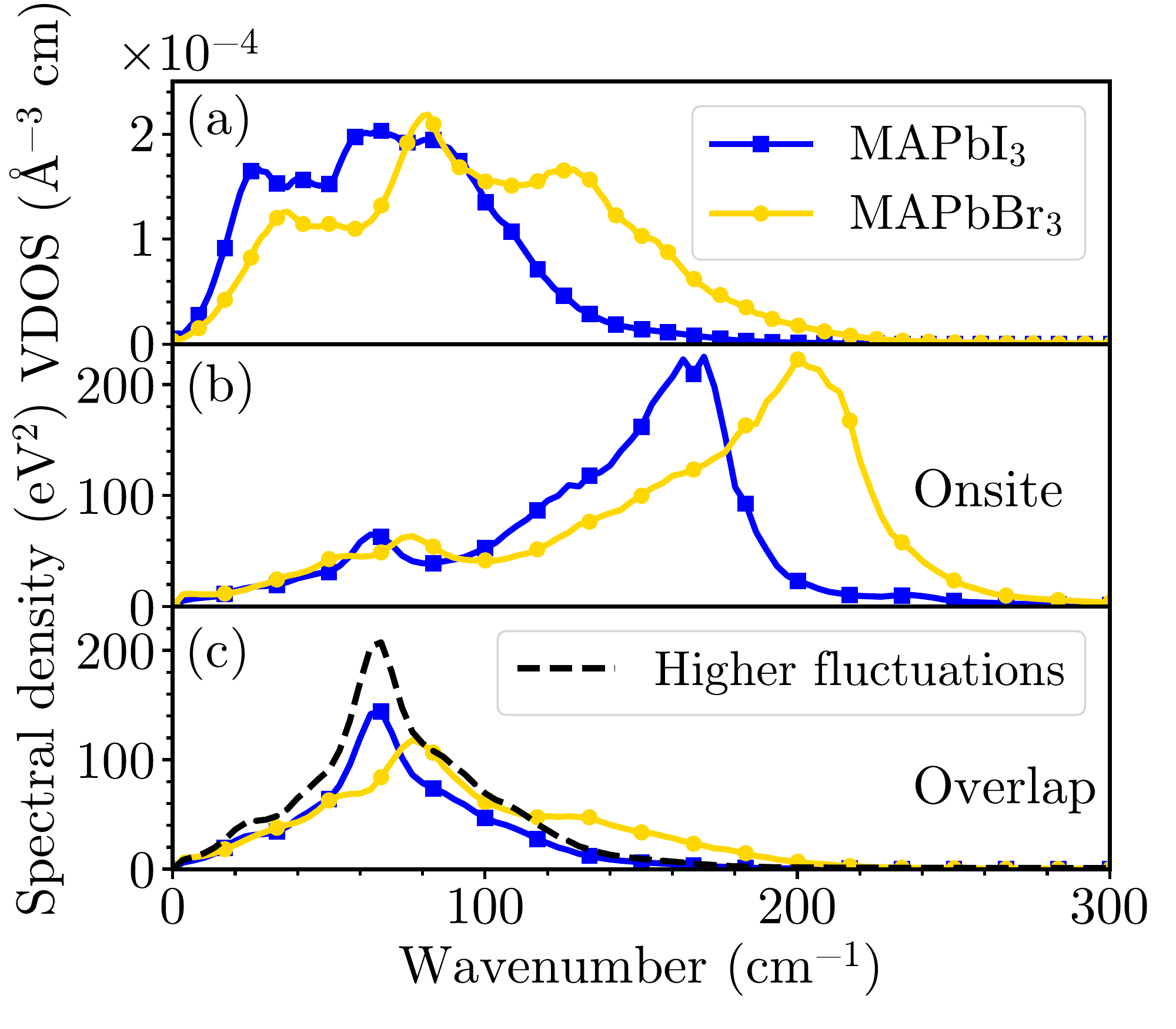}
    \caption{Vibrational density of states (VDOS, panel a) and spectral densities of onsite (panel b) and the $t_\mathrm{pp\sigma}$ overlap-terms (panel c) at 300\,K.
    Spectral densities were computed from instantaneous fluctuations occurring in the multi-scale TB model.
    The dashed lines shows an artificial signal for the $t_\mathrm{pp\sigma}$ spectral density where fluctuations have been manually increased by 20\,\%, which caused a mobility reduction of ${\approx}20\mathrm{cm}^2$/Vs.
    All panels show the low-frequency region of the spectra.}
    \label{fig:3}
\end{figure}%
We investigate the connections between anharmonicity and dynamic disorder through a statistical analysis of the temperature-dependent atomic and electronic dynamics.
The vibrational density of states (VDOS) at 300\,K (see Fig.~\ref{fig:3}a) shows prominent THz-range contributions in both compounds and a slight shift of the MAPbBr$_3$ spectrum to higher frequencies. A spectral analysis of the finite-temperature fluctuations of the corresponding onsite and overlap terms in the TB model is presented in Figs.~\ref{fig:3}b and c.
Importantly, pronounced intensities in the $t_\mathrm{pp\sigma}$ overlap fluctuations, which is the dominant scattering channel for carriers in these materials \cite{SM, mayers_HowLatticeCharge_2018}, appear in a similarly low frequency region as the pronounced intensities in the VDOS (\textit{cf.} Figs.~\ref{fig:3}a and c).
Furthermore, a shift to higher frequencies is seen in the $t_\mathrm{pp\sigma}$ fluctuations for MAPbBr$_3$, similar to what is observed in the VDOS.
Therefore, the overlap fluctuations follow the VDOS in both compounds.
At 300\,K, these fluctuations are more pronounced in the more anharmonic MAPbBr$_3$:
standard deviations of the $t_\mathrm{pp\sigma}$ fluctuations are 0.21\,eV and 0.23\,eV for MAPbI$_3$ and MAPbBr$_3$ at 300\,K, respectively, confirming that the more anharmonic MAPbBr$_3$ is more dynamically disordered at that temperature.\\
To connect these findings to  carrier dynamics, we construct a series of artificial onsite and coupling signals augmenting the original TB Hamiltonian (see Ref. \cite{SM}).
Interestingly, when we increase the fluctuations in the $t_\mathrm{pp\sigma}$ couplings of MAPbI$_3$ (see Fig.~\ref{fig:3}c for the corresponding spectral density) its carrier mobility is significantly reduced (by 20\,cm$^2$/Vs) at 300\,K, while changes to the onsite terms have a smaller effect and a shift of the fluctuations to higher frequencies is inconsequential \cite{SM}.
Therefore, what distinguishes the carrier dynamics in the two materials at lower temperature are differences in the degree of dynamic disorder.\\
Having established the critical role of dynamic disorder for the carrier mobility through the $t_\mathrm{pp\sigma}$ fluctuations, it is interesting to analyze their temperature dependencies in both materials.
Fig.~\ref{fig:4} shows temperature-dependent relative fluctuations in Pb-\textit{X} bond distances and
$t_\mathrm{pp\sigma}$ overlaps for the two compounds.
Concurrent with the more anharmonic behavior of MAPbBr$_3$ at lower temperatures are larger relative fluctuations in bond distances compared to MAPbI$_3$, which become more similar as temperature is raised.
Similarly, the relative fluctuations in $t_\mathrm{pp\sigma}$ overlaps are larger in MAPbBr$_3$ at lower temperatures, but those of MAPbI$_3$ increase more strongly as temperature is raised, until they are very similar in the two materials at 350\,K where both materials are in the cubic phase.
Together with the findings outlined above, these data show that anharmonicity and dynamic disorder are {\em microscopically connected} and appear to be the two critical factors determining the carrier mobility and its temperature dependence in HaPs.\\
Finally, we discuss the implications of our findings for modeling of electron-phonon interactions in soft, anharmonic materials more broadly. It is useful to attempt to rationalize our findings presented in Fig.~\ref{fig:2} from a purely electronic structure perspective, using the static crystal structures and effective masses of the tetragonal and cubic phase for both materials.
We find that the effective mass of MAPbI$_3$ is indeed lower than that of MAPbBr$_3$, in line with the majority of previous studies\cite{chang_FirstprinciplesStudyStructural_2004, feng_CrystalStructuresOptical_2014, chen_SpotlightOrganicInorganic_2015, galkowski_DeterminationExcitonBinding_2016, mosconi_ElectronicOpticalProperties_2016, jong_InfluenceHalideComposition_2016, egger_WhatRemainsUnexplained_2018a, zu_ConstructingElectronicStructure_2019a, yamada_PolaronMassesCH_2021, traore_BandGapEffective_2022}, 
which seemingly explains the trend we find up to ${\approx}300$\,K.
However, changes in the \textit{relative differences} of the effective masses of the two compounds upon undergoing the tetragonal-to-cubic phase transition show that they are significantly more similar in the tetragonal phase than in the cubic phase \cite{SM}, which is opposite to the trend exposed in Fig.~\ref{fig:2}.\\
The effective masses of the compounds alone cannot explain our findings, which signifies potential limitations of a momentum-space quasiparticle representation of carriers at finite temperature in these systems.  Indeed, established electron-phonon models rooted in band theory apply static electronic band structures as a starting point in a perturbative treatment of finite-temperature effects, for which the aforementioned findings concerning effective masses suggest that their predictive power may be limited. As a case in point, previous theoretical studies applying band theory and the Boltzmann transport equation generically report \textit{stronger} mobility temperature dependencies at lower temperatures when HaPs adopt lower-symmetry phases \cite{filippetti_LowElectronpolarOptical_2016a, ponce_OriginLowCarrier_2019a, irvine_QuantifyingPolaronicEffects_2021}.
By contrast, experimental studies on various HaP compounds have consistently reported stronger mobility temperature dependencies in high-symmetry phases \cite{karakus_PhononElectronScattering_2015a, yi_IntrinsicChargeTransport_2016a,  shrestha_AssessingTemperatureDependence_2018a, biewald_TemperatureDependentAmbipolarCharge_2019a}.\\
\hl{In contrast to methods based on a momentum-space representation and perturbative electron-phonon couplings, the approach adopted here does not rely on band theory. 
Rather it includes all carrier-phonon scattering effects that arise in the semiclassical treatment of the finite-temperature atomic motion in the material}~\cite{mayers_HowLatticeCharge_2018,schilcher_SignificancePolaronsDynamic_2021}.
\hl{Several methods based on effective harmonic potentials have been developed to extend the perturbative momentum-space electron-phonon methods to anharmonic materials}~\cite{hellman_TemperatureDependentEffective_2013, zacharias_AnharmonicElectronphononCoupling_2023}. 
\hl{However, the non-Gaussian nature of the atomic displacements in Fig.}~\ref{fig:1}\hl{ suggests that no effective harmonic potential would fully capture the lattice dynamics in MAPbI$_3$ and MAPbBr$_3$. 
This has been discussed also for the related CsPbBr$_3$ compound in previous work} \cite{klarbring_AnharmonicityUltralowThermal_2020}.
\hl{Additionally, perturbative momentum-space methods typically employ a linear electron-phonon coupling and neglect phonon scattering resulting from anharmonicity, which are central to the lattice dynamics in HaPs}~\cite{beecher_DirectObservationDynamic_2016, yaffe_LocalPolarFluctuations_2017a, wu_LightinducedPicosecondRotational_2017, yang_SpontaneousOctahedralTilting_2017, gold-parker_AcousticPhononLifetimes_2018a, klarbring_LowenergyPathsOctahedral_2019, liu_ThermalDisorderBond_2019, ferreira_DirectEvidenceWeakly_2020, lanigan-atkins_TwodimensionalOverdampedFluctuations_2021, debnath_CoherentVibrationalDynamics_2021, cannelli_AtomicLevelDescriptionThermal_2022, zhu_ProbingDisorderCubic_2022, fransson_LimitsPhononQuasiparticle_2023, weadock_NatureDynamicLocal_2023a};
\hl{such limitations are not present in our method. 
Moreover, our treatment of the quantum dynamics also accounts for the electronic dynamics at a higher level of theory than semi-classical Boltzmann transport approaches and it naturally includes interband effects which are often excluded in other methods.}
\hl{Therefore}, we hypothesize that increased anharmonicity and dynamic disorder \hl{we found using this method} reduces carrier mobilities and thus resolves \hl{remaining contradictions between experiment and theory on the carrier-scattering mechanisms that are active in HaPs}.  
Hence, including finite-temperature effects directly into a dynamic disorder-based representation of the carrier scattering enhances the predictive power of the theory.\\
In summary, we have studied the connections betweeen anharmonicity and dynamic disorder by comparing two prototypical variants of anharmonic semiconductors, namely the HaPs MAPbI$_3$ and MAPbBr$_3$.
Using a TB model that is parameterized from first-principles calculations, MD simulations,  and semiclassical quantum dynamics, we have rationalized subtle differences in the power-law behavior of temperature-dependent mobilities for both compounds.  
Most critically, we demonstrated that charge carriers follow the atomic dynamics by revealing that in the temperature region where MAPbBr$_3$ is more anharmonic, its charge carrier mobility is reduced.
Our model and the real-space picture underlying it enabled us to determine that anharmonicity and dynamic disorder are connected to one another, and that they critically impact carrier mobility characteristics, including their temperature dependencies, in a systematic manner.
These findings have relevance for development of structure-property relations  which promises to be useful for tuning the properties of a wide class of semiconductors and anharmonic solids, as well as for the devices which utilize the relations for materials design.

 \begin{acknowledgments}
\noindent We thank Andrew M. Rappe for past collaborations on related work.
Funding provided by the Alexander von Humboldt-Foundation in the framework of the Sofja Kovalevskaja Award, endowed by the German Federal Ministry of Education and Research, by the Deutsche Forschungsgemeinschaft (DFG, German Research Foundation) \textit{via} Germany's Excellence Strategy - EXC 2089/1-390776260, and by TU Munich - IAS, funded by the German Excellence Initiative and the European Union Seventh Framework Programme under Grant Agreement No. 291763, are gratefully acknowledged.
The work of DRR was performed with support from the U.S. Department of Energy, Office of Science, Office of Advanced Scientific Computing Research, Scientific Discovery through Advanced Computing (SciDAC) program, under Award No. DE-SC0022088.
This work was supported by the user program of the Molecular Foundry, a DOE Office of Science User Facility supported by the Office of Science of the U.S. Department of Energy under Contract No. DE-AC02-05CH11231.
The Gauss Centre for Supercomputing e.V. is acknowledged for providing computing time through the John von Neumann Institute for Computing on the GCS Supercomputer JUWELS at J\"ulich Supercomputing Centre. This research used resources of the National Energy Research Scientific Computing Center, a DOE Office of Science User Facility supported by the Office of Science of the U.S. Department of Energy under Contract No. DE-AC02-05CH11231.
\end{acknowledgments}

\bibliographystyle{apsrev4-2}

\bibliography{literature.bib}

\begin{thebibliography}{83}%
\makeatletter
\providecommand \@ifxundefined [1]{%
 \@ifx{#1\undefined}
}%
\providecommand \@ifnum [1]{%
 \ifnum #1\expandafter \@firstoftwo
 \else \expandafter \@secondoftwo
 \fi
}%
\providecommand \@ifx [1]{%
 \ifx #1\expandafter \@firstoftwo
 \else \expandafter \@secondoftwo
 \fi
}%
\providecommand \natexlab [1]{#1}%
\providecommand \enquote  [1]{``#1''}%
\providecommand \bibnamefont  [1]{#1}%
\providecommand \bibfnamefont [1]{#1}%
\providecommand \citenamefont [1]{#1}%
\providecommand \href@noop [0]{\@secondoftwo}%
\providecommand \href [0]{\begingroup \@sanitize@url \@href}%
\providecommand \@href[1]{\@@startlink{#1}\@@href}%
\providecommand \@@href[1]{\endgroup#1\@@endlink}%
\providecommand \@sanitize@url [0]{\catcode `\\12\catcode `\$12\catcode
  `\&12\catcode `\#12\catcode `\^12\catcode `\_12\catcode `\%12\relax}%
\providecommand \@@startlink[1]{}%
\providecommand \@@endlink[0]{}%
\providecommand \url  [0]{\begingroup\@sanitize@url \@url }%
\providecommand \@url [1]{\endgroup\@href {#1}{\urlprefix }}%
\providecommand \urlprefix  [0]{URL }%
\providecommand \Eprint [0]{\href }%
\providecommand \doibase [0]{https://doi.org/}%
\providecommand \selectlanguage [0]{\@gobble}%
\providecommand \bibinfo  [0]{\@secondoftwo}%
\providecommand \bibfield  [0]{\@secondoftwo}%
\providecommand \translation [1]{[#1]}%
\providecommand \BibitemOpen [0]{}%
\providecommand \bibitemStop [0]{}%
\providecommand \bibitemNoStop [0]{.\EOS\space}%
\providecommand \EOS [0]{\spacefactor3000\relax}%
\providecommand \BibitemShut  [1]{\csname bibitem#1\endcsname}%
\let\auto@bib@innerbib\@empty
\bibitem [{\citenamefont {Snaith}(2013)}]{snaith_PerovskitesEmergenceNew_2013}%
  \BibitemOpen
  \bibfield  {author} {\bibinfo {author} {\bibfnamefont {H.~J.}\ \bibnamefont
  {Snaith}},\ }\href {https://doi.org/10.1021/jz4020162} {\bibfield  {journal}
  {\bibinfo  {journal} {J. Phys. Chem. Lett.}\ }\textbf {\bibinfo {volume}
  {4}},\ \bibinfo {pages} {3623} (\bibinfo {year} {2013})}\BibitemShut
  {NoStop}%
\bibitem [{\citenamefont {Green}\ \emph {et~al.}(2014)\citenamefont {Green},
  \citenamefont {{Ho-Baillie}},\ and\ \citenamefont
  {Snaith}}]{green_EmergencePerovskiteSolar_2014}%
  \BibitemOpen
  \bibfield  {author} {\bibinfo {author} {\bibfnamefont {M.~A.}\ \bibnamefont
  {Green}}, \bibinfo {author} {\bibfnamefont {A.}~\bibnamefont
  {{Ho-Baillie}}},\ and\ \bibinfo {author} {\bibfnamefont {H.~J.}\ \bibnamefont
  {Snaith}},\ }\href {https://doi.org/10.1038/nphoton.2014.134} {\bibfield
  {journal} {\bibinfo  {journal} {Nat. Photon.}\ }\textbf {\bibinfo {volume}
  {8}},\ \bibinfo {pages} {506} (\bibinfo {year} {2014})}\BibitemShut {NoStop}%
\bibitem [{\citenamefont {Stranks}\ and\ \citenamefont
  {Snaith}(2015)}]{stranks_MetalhalidePerovskitesPhotovoltaic_2015}%
  \BibitemOpen
  \bibfield  {author} {\bibinfo {author} {\bibfnamefont {S.~D.}\ \bibnamefont
  {Stranks}}\ and\ \bibinfo {author} {\bibfnamefont {H.~J.}\ \bibnamefont
  {Snaith}},\ }\href {https://doi.org/10.1038/nnano.2015.90} {\bibfield
  {journal} {\bibinfo  {journal} {Nat. Nanotechnol.}\ }\textbf {\bibinfo
  {volume} {10}},\ \bibinfo {pages} {391} (\bibinfo {year} {2015})}\BibitemShut
  {NoStop}%
\bibitem [{\citenamefont {{Correa-Baena}}\ \emph {et~al.}(2017)\citenamefont
  {{Correa-Baena}}, \citenamefont {Saliba}, \citenamefont {Buonassisi},
  \citenamefont {Gr{\"a}tzel}, \citenamefont {Abate}, \citenamefont {Tress},\
  and\ \citenamefont
  {Hagfeldt}}]{correa-baena_PromisesChallengesPerovskite_2017}%
  \BibitemOpen
  \bibfield  {author} {\bibinfo {author} {\bibfnamefont {J.-P.}\ \bibnamefont
  {{Correa-Baena}}}, \bibinfo {author} {\bibfnamefont {M.}~\bibnamefont
  {Saliba}}, \bibinfo {author} {\bibfnamefont {T.}~\bibnamefont {Buonassisi}},
  \bibinfo {author} {\bibfnamefont {M.}~\bibnamefont {Gr{\"a}tzel}}, \bibinfo
  {author} {\bibfnamefont {A.}~\bibnamefont {Abate}}, \bibinfo {author}
  {\bibfnamefont {W.}~\bibnamefont {Tress}},\ and\ \bibinfo {author}
  {\bibfnamefont {A.}~\bibnamefont {Hagfeldt}},\ }\href
  {https://doi.org/10.1126/science.aam6323} {\bibfield  {journal} {\bibinfo
  {journal} {Science}\ }\textbf {\bibinfo {volume} {358}},\ \bibinfo {pages}
  {739} (\bibinfo {year} {2017})}\BibitemShut {NoStop}%
\bibitem [{\citenamefont {Li}\ \emph {et~al.}(2017)\citenamefont {Li},
  \citenamefont {Wang}, \citenamefont {Deschler}, \citenamefont {Gao},
  \citenamefont {Friend},\ and\ \citenamefont
  {Cheetham}}]{li_ChemicallyDiverseMultifunctional_2017}%
  \BibitemOpen
  \bibfield  {author} {\bibinfo {author} {\bibfnamefont {W.}~\bibnamefont
  {Li}}, \bibinfo {author} {\bibfnamefont {Z.}~\bibnamefont {Wang}}, \bibinfo
  {author} {\bibfnamefont {F.}~\bibnamefont {Deschler}}, \bibinfo {author}
  {\bibfnamefont {S.}~\bibnamefont {Gao}}, \bibinfo {author} {\bibfnamefont
  {R.~H.}\ \bibnamefont {Friend}},\ and\ \bibinfo {author} {\bibfnamefont
  {A.~K.}\ \bibnamefont {Cheetham}},\ }\href
  {https://doi.org/10.1038/natrevmats.2016.99} {\bibfield  {journal} {\bibinfo
  {journal} {Nat. Rev. Mater.}\ }\textbf {\bibinfo {volume} {2}},\ \bibinfo
  {pages} {16099} (\bibinfo {year} {2017})}\BibitemShut {NoStop}%
\bibitem [{\citenamefont {Nayak}\ \emph {et~al.}(2019)\citenamefont {Nayak},
  \citenamefont {Mahesh}, \citenamefont {Snaith},\ and\ \citenamefont
  {Cahen}}]{nayak_PhotovoltaicSolarCell_2019}%
  \BibitemOpen
  \bibfield  {author} {\bibinfo {author} {\bibfnamefont {P.~K.}\ \bibnamefont
  {Nayak}}, \bibinfo {author} {\bibfnamefont {S.}~\bibnamefont {Mahesh}},
  \bibinfo {author} {\bibfnamefont {H.~J.}\ \bibnamefont {Snaith}},\ and\
  \bibinfo {author} {\bibfnamefont {D.}~\bibnamefont {Cahen}},\ }\href
  {https://doi.org/10.1038/s41578-019-0097-0} {\bibfield  {journal} {\bibinfo
  {journal} {Nat. Rev. Mater.}\ }\textbf {\bibinfo {volume} {4}},\ \bibinfo
  {pages} {269} (\bibinfo {year} {2019})}\BibitemShut {NoStop}%
\bibitem [{\citenamefont {Brenner}\ \emph {et~al.}(2016)\citenamefont
  {Brenner}, \citenamefont {Egger}, \citenamefont {Kronik}, \citenamefont
  {Hodes},\ and\ \citenamefont {Cahen}}]{brenner_HybridOrganicInorganic_2016}%
  \BibitemOpen
  \bibfield  {author} {\bibinfo {author} {\bibfnamefont {T.~M.}\ \bibnamefont
  {Brenner}}, \bibinfo {author} {\bibfnamefont {D.~A.}\ \bibnamefont {Egger}},
  \bibinfo {author} {\bibfnamefont {L.}~\bibnamefont {Kronik}}, \bibinfo
  {author} {\bibfnamefont {G.}~\bibnamefont {Hodes}},\ and\ \bibinfo {author}
  {\bibfnamefont {D.}~\bibnamefont {Cahen}},\ }\href
  {https://doi.org/10.1038/natrevmats.2015.7} {\bibfield  {journal} {\bibinfo
  {journal} {Nat. Rev. Mater.}\ }\textbf {\bibinfo {volume} {1}},\ \bibinfo
  {pages} {15007} (\bibinfo {year} {2016})}\BibitemShut {NoStop}%
\bibitem [{\citenamefont {Stoumpos}\ and\ \citenamefont
  {Kanatzidis}(2016)}]{stoumpos_HalidePerovskitesPoor_2016a}%
  \BibitemOpen
  \bibfield  {author} {\bibinfo {author} {\bibfnamefont {C.~C.}\ \bibnamefont
  {Stoumpos}}\ and\ \bibinfo {author} {\bibfnamefont {M.~G.}\ \bibnamefont
  {Kanatzidis}},\ }\href {https://doi.org/10.1002/adma.201600265} {\bibfield
  {journal} {\bibinfo  {journal} {Adv. Mater.}\ }\textbf {\bibinfo {volume}
  {28}},\ \bibinfo {pages} {5778} (\bibinfo {year} {2016})}\BibitemShut
  {NoStop}%
\bibitem [{\citenamefont {Stranks}\ \emph {et~al.}(2013)\citenamefont
  {Stranks}, \citenamefont {Eperon}, \citenamefont {Grancini}, \citenamefont
  {Menelaou}, \citenamefont {Alcocer}, \citenamefont {Leijtens}, \citenamefont
  {Herz}, \citenamefont {Petrozza},\ and\ \citenamefont
  {Snaith}}]{stranks_ElectronHoleDiffusionLengths_2013}%
  \BibitemOpen
  \bibfield  {author} {\bibinfo {author} {\bibfnamefont {S.~D.}\ \bibnamefont
  {Stranks}}, \bibinfo {author} {\bibfnamefont {G.~E.}\ \bibnamefont {Eperon}},
  \bibinfo {author} {\bibfnamefont {G.}~\bibnamefont {Grancini}}, \bibinfo
  {author} {\bibfnamefont {C.}~\bibnamefont {Menelaou}}, \bibinfo {author}
  {\bibfnamefont {M.~J.~P.}\ \bibnamefont {Alcocer}}, \bibinfo {author}
  {\bibfnamefont {T.}~\bibnamefont {Leijtens}}, \bibinfo {author}
  {\bibfnamefont {L.~M.}\ \bibnamefont {Herz}}, \bibinfo {author}
  {\bibfnamefont {A.}~\bibnamefont {Petrozza}},\ and\ \bibinfo {author}
  {\bibfnamefont {H.~J.}\ \bibnamefont {Snaith}},\ }\href
  {https://doi.org/10.1126/science.1243982} {\bibfield  {journal} {\bibinfo
  {journal} {Science}\ }\textbf {\bibinfo {volume} {342}},\ \bibinfo {pages}
  {341} (\bibinfo {year} {2013})}\BibitemShut {NoStop}%
\bibitem [{\citenamefont {Saparov}\ and\ \citenamefont
  {Mitzi}(2016)}]{saparov_OrganicInorganicPerovskites_2016}%
  \BibitemOpen
  \bibfield  {author} {\bibinfo {author} {\bibfnamefont {B.}~\bibnamefont
  {Saparov}}\ and\ \bibinfo {author} {\bibfnamefont {D.~B.}\ \bibnamefont
  {Mitzi}},\ }\href {https://doi.org/10.1021/acs.chemrev.5b00715} {\bibfield
  {journal} {\bibinfo  {journal} {Chem. Rev.}\ }\textbf {\bibinfo {volume}
  {116}},\ \bibinfo {pages} {4558} (\bibinfo {year} {2016})}\BibitemShut
  {NoStop}%
\bibitem [{\citenamefont {Lal}\ \emph {et~al.}(2017)\citenamefont {Lal},
  \citenamefont {Dkhissi}, \citenamefont {Li}, \citenamefont {Hou},
  \citenamefont {Cheng},\ and\ \citenamefont
  {Bach}}]{lal_PerovskiteTandemSolar_2017}%
  \BibitemOpen
  \bibfield  {author} {\bibinfo {author} {\bibfnamefont {N.~N.}\ \bibnamefont
  {Lal}}, \bibinfo {author} {\bibfnamefont {Y.}~\bibnamefont {Dkhissi}},
  \bibinfo {author} {\bibfnamefont {W.}~\bibnamefont {Li}}, \bibinfo {author}
  {\bibfnamefont {Q.}~\bibnamefont {Hou}}, \bibinfo {author} {\bibfnamefont
  {Y.-B.}\ \bibnamefont {Cheng}},\ and\ \bibinfo {author} {\bibfnamefont
  {U.}~\bibnamefont {Bach}},\ }\href {https://doi.org/10.1002/aenm.201602761}
  {\bibfield  {journal} {\bibinfo  {journal} {Adv. Energy Mater.}\ }\textbf
  {\bibinfo {volume} {7}},\ \bibinfo {pages} {1602761} (\bibinfo {year}
  {2017})}\BibitemShut {NoStop}%
\bibitem [{\citenamefont {Egger}\ \emph {et~al.}(2018)\citenamefont {Egger},
  \citenamefont {Bera}, \citenamefont {Cahen}, \citenamefont {Hodes},
  \citenamefont {Kirchartz}, \citenamefont {Kronik}, \citenamefont {Lovrincic},
  \citenamefont {Rappe}, \citenamefont {Reichman},\ and\ \citenamefont
  {Yaffe}}]{egger_WhatRemainsUnexplained_2018a}%
  \BibitemOpen
  \bibfield  {author} {\bibinfo {author} {\bibfnamefont {D.~A.}\ \bibnamefont
  {Egger}}, \bibinfo {author} {\bibfnamefont {A.}~\bibnamefont {Bera}},
  \bibinfo {author} {\bibfnamefont {D.}~\bibnamefont {Cahen}}, \bibinfo
  {author} {\bibfnamefont {G.}~\bibnamefont {Hodes}}, \bibinfo {author}
  {\bibfnamefont {T.}~\bibnamefont {Kirchartz}}, \bibinfo {author}
  {\bibfnamefont {L.}~\bibnamefont {Kronik}}, \bibinfo {author} {\bibfnamefont
  {R.}~\bibnamefont {Lovrincic}}, \bibinfo {author} {\bibfnamefont {A.~M.}\
  \bibnamefont {Rappe}}, \bibinfo {author} {\bibfnamefont {D.~R.}\ \bibnamefont
  {Reichman}},\ and\ \bibinfo {author} {\bibfnamefont {O.}~\bibnamefont
  {Yaffe}},\ }\href {https://doi.org/10.1002/adma.201800691} {\bibfield
  {journal} {\bibinfo  {journal} {Adv. Mater.}\ }\textbf {\bibinfo {volume}
  {30}},\ \bibinfo {pages} {1800691} (\bibinfo {year} {2018})}\BibitemShut
  {NoStop}%
\bibitem [{\citenamefont
  {Herz}(2017)}]{herz_ChargeCarrierMobilitiesMetal_2017a}%
  \BibitemOpen
  \bibfield  {author} {\bibinfo {author} {\bibfnamefont {L.~M.}\ \bibnamefont
  {Herz}},\ }\href {https://doi.org/10.1021/acsenergylett.7b00276} {\bibfield
  {journal} {\bibinfo  {journal} {ACS Energy Lett.}\ }\textbf {\bibinfo
  {volume} {2}},\ \bibinfo {pages} {1539} (\bibinfo {year} {2017})}\BibitemShut
  {NoStop}%
\bibitem [{\citenamefont {Schilcher}\ \emph {et~al.}(2021)\citenamefont
  {Schilcher}, \citenamefont {Robinson}, \citenamefont {Abramovitch},
  \citenamefont {Tan}, \citenamefont {Rappe}, \citenamefont {Reichman},\ and\
  \citenamefont {Egger}}]{schilcher_SignificancePolaronsDynamic_2021}%
  \BibitemOpen
  \bibfield  {author} {\bibinfo {author} {\bibfnamefont {M.~J.}\ \bibnamefont
  {Schilcher}}, \bibinfo {author} {\bibfnamefont {P.~J.}\ \bibnamefont
  {Robinson}}, \bibinfo {author} {\bibfnamefont {D.~J.}\ \bibnamefont
  {Abramovitch}}, \bibinfo {author} {\bibfnamefont {L.~Z.}\ \bibnamefont
  {Tan}}, \bibinfo {author} {\bibfnamefont {A.~M.}\ \bibnamefont {Rappe}},
  \bibinfo {author} {\bibfnamefont {D.~R.}\ \bibnamefont {Reichman}},\ and\
  \bibinfo {author} {\bibfnamefont {D.~A.}\ \bibnamefont {Egger}},\ }\href
  {https://doi.org/10.1021/acsenergylett.1c00506} {\bibfield  {journal}
  {\bibinfo  {journal} {ACS Energy Lett.}\ }\textbf {\bibinfo {volume} {6}},\
  \bibinfo {pages} {2162} (\bibinfo {year} {2021})}\BibitemShut {NoStop}%
\bibitem [{\citenamefont {Beecher}\ \emph {et~al.}(2016)\citenamefont
  {Beecher}, \citenamefont {Semonin}, \citenamefont {Skelton}, \citenamefont
  {Frost}, \citenamefont {Terban}, \citenamefont {Zhai}, \citenamefont
  {Alatas}, \citenamefont {Owen}, \citenamefont {Walsh},\ and\ \citenamefont
  {Billinge}}]{beecher_DirectObservationDynamic_2016}%
  \BibitemOpen
  \bibfield  {author} {\bibinfo {author} {\bibfnamefont {A.~N.}\ \bibnamefont
  {Beecher}}, \bibinfo {author} {\bibfnamefont {O.~E.}\ \bibnamefont
  {Semonin}}, \bibinfo {author} {\bibfnamefont {J.~M.}\ \bibnamefont
  {Skelton}}, \bibinfo {author} {\bibfnamefont {J.~M.}\ \bibnamefont {Frost}},
  \bibinfo {author} {\bibfnamefont {M.~W.}\ \bibnamefont {Terban}}, \bibinfo
  {author} {\bibfnamefont {H.}~\bibnamefont {Zhai}}, \bibinfo {author}
  {\bibfnamefont {A.}~\bibnamefont {Alatas}}, \bibinfo {author} {\bibfnamefont
  {J.~S.}\ \bibnamefont {Owen}}, \bibinfo {author} {\bibfnamefont
  {A.}~\bibnamefont {Walsh}},\ and\ \bibinfo {author} {\bibfnamefont
  {S.~J.~L.}\ \bibnamefont {Billinge}},\ }\href
  {https://doi.org/10.1021/acsenergylett.6b00381} {\bibfield  {journal}
  {\bibinfo  {journal} {ACS Energy Lett.}\ }\textbf {\bibinfo {volume} {1}},\
  \bibinfo {pages} {880} (\bibinfo {year} {2016})}\BibitemShut {NoStop}%
\bibitem [{\citenamefont {Yaffe}\ \emph {et~al.}(2017)\citenamefont {Yaffe},
  \citenamefont {Guo}, \citenamefont {Tan}, \citenamefont {Egger},
  \citenamefont {Hull}, \citenamefont {Stoumpos}, \citenamefont {Zheng},
  \citenamefont {Heinz}, \citenamefont {Kronik}, \citenamefont {Kanatzidis},
  \citenamefont {Owen}, \citenamefont {Rappe}, \citenamefont {Pimenta},\ and\
  \citenamefont {Brus}}]{yaffe_LocalPolarFluctuations_2017a}%
  \BibitemOpen
  \bibfield  {author} {\bibinfo {author} {\bibfnamefont {O.}~\bibnamefont
  {Yaffe}}, \bibinfo {author} {\bibfnamefont {Y.}~\bibnamefont {Guo}}, \bibinfo
  {author} {\bibfnamefont {L.~Z.}\ \bibnamefont {Tan}}, \bibinfo {author}
  {\bibfnamefont {D.~A.}\ \bibnamefont {Egger}}, \bibinfo {author}
  {\bibfnamefont {T.}~\bibnamefont {Hull}}, \bibinfo {author} {\bibfnamefont
  {C.~C.}\ \bibnamefont {Stoumpos}}, \bibinfo {author} {\bibfnamefont
  {F.}~\bibnamefont {Zheng}}, \bibinfo {author} {\bibfnamefont {T.~F.}\
  \bibnamefont {Heinz}}, \bibinfo {author} {\bibfnamefont {L.}~\bibnamefont
  {Kronik}}, \bibinfo {author} {\bibfnamefont {M.~G.}\ \bibnamefont
  {Kanatzidis}}, \bibinfo {author} {\bibfnamefont {J.~S.}\ \bibnamefont
  {Owen}}, \bibinfo {author} {\bibfnamefont {A.~M.}\ \bibnamefont {Rappe}},
  \bibinfo {author} {\bibfnamefont {M.~A.}\ \bibnamefont {Pimenta}},\ and\
  \bibinfo {author} {\bibfnamefont {L.~E.}\ \bibnamefont {Brus}},\ }\href
  {https://doi.org/10.1103/PhysRevLett.118.136001} {\bibfield  {journal}
  {\bibinfo  {journal} {Phys. Rev. Lett.}\ }\textbf {\bibinfo {volume} {118}},\
  \bibinfo {pages} {136001} (\bibinfo {year} {2017})}\BibitemShut {NoStop}%
\bibitem [{\citenamefont {Wu}\ \emph {et~al.}(2017)\citenamefont {Wu},
  \citenamefont {Tan}, \citenamefont {Shen}, \citenamefont {Hu}, \citenamefont
  {Miyata}, \citenamefont {Trinh}, \citenamefont {Li}, \citenamefont {Coffee},
  \citenamefont {Liu}, \citenamefont {Egger}, \citenamefont {Makasyuk},
  \citenamefont {Zheng}, \citenamefont {Fry}, \citenamefont {Robinson},
  \citenamefont {Smith}, \citenamefont {Guzelturk}, \citenamefont {Karunadasa},
  \citenamefont {Wang}, \citenamefont {Zhu}, \citenamefont {Kronik},
  \citenamefont {Rappe},\ and\ \citenamefont
  {Lindenberg}}]{wu_LightinducedPicosecondRotational_2017}%
  \BibitemOpen
  \bibfield  {author} {\bibinfo {author} {\bibfnamefont {X.}~\bibnamefont
  {Wu}}, \bibinfo {author} {\bibfnamefont {L.~Z.}\ \bibnamefont {Tan}},
  \bibinfo {author} {\bibfnamefont {X.}~\bibnamefont {Shen}}, \bibinfo {author}
  {\bibfnamefont {T.}~\bibnamefont {Hu}}, \bibinfo {author} {\bibfnamefont
  {K.}~\bibnamefont {Miyata}}, \bibinfo {author} {\bibfnamefont {M.~T.}\
  \bibnamefont {Trinh}}, \bibinfo {author} {\bibfnamefont {R.}~\bibnamefont
  {Li}}, \bibinfo {author} {\bibfnamefont {R.}~\bibnamefont {Coffee}}, \bibinfo
  {author} {\bibfnamefont {S.}~\bibnamefont {Liu}}, \bibinfo {author}
  {\bibfnamefont {D.~A.}\ \bibnamefont {Egger}}, \bibinfo {author}
  {\bibfnamefont {I.}~\bibnamefont {Makasyuk}}, \bibinfo {author}
  {\bibfnamefont {Q.}~\bibnamefont {Zheng}}, \bibinfo {author} {\bibfnamefont
  {A.}~\bibnamefont {Fry}}, \bibinfo {author} {\bibfnamefont {J.~S.}\
  \bibnamefont {Robinson}}, \bibinfo {author} {\bibfnamefont {M.~D.}\
  \bibnamefont {Smith}}, \bibinfo {author} {\bibfnamefont {B.}~\bibnamefont
  {Guzelturk}}, \bibinfo {author} {\bibfnamefont {H.~I.}\ \bibnamefont
  {Karunadasa}}, \bibinfo {author} {\bibfnamefont {X.}~\bibnamefont {Wang}},
  \bibinfo {author} {\bibfnamefont {X.}~\bibnamefont {Zhu}}, \bibinfo {author}
  {\bibfnamefont {L.}~\bibnamefont {Kronik}}, \bibinfo {author} {\bibfnamefont
  {A.~M.}\ \bibnamefont {Rappe}},\ and\ \bibinfo {author} {\bibfnamefont
  {A.~M.}\ \bibnamefont {Lindenberg}},\ }\href
  {https://doi.org/10.1126/sciadv.1602388} {\bibfield  {journal} {\bibinfo
  {journal} {Sci. Adv.}\ }\textbf {\bibinfo {volume} {3}},\ \bibinfo {pages}
  {e1602388} (\bibinfo {year} {2017})}\BibitemShut {NoStop}%
\bibitem [{\citenamefont {Yang}\ \emph {et~al.}(2017)\citenamefont {Yang},
  \citenamefont {Skelton}, \citenamefont {{da Silva}}, \citenamefont {Frost},\
  and\ \citenamefont {Walsh}}]{yang_SpontaneousOctahedralTilting_2017}%
  \BibitemOpen
  \bibfield  {author} {\bibinfo {author} {\bibfnamefont {R.~X.}\ \bibnamefont
  {Yang}}, \bibinfo {author} {\bibfnamefont {J.~M.}\ \bibnamefont {Skelton}},
  \bibinfo {author} {\bibfnamefont {E.~L.}\ \bibnamefont {{da Silva}}},
  \bibinfo {author} {\bibfnamefont {J.~M.}\ \bibnamefont {Frost}},\ and\
  \bibinfo {author} {\bibfnamefont {A.}~\bibnamefont {Walsh}},\ }\href
  {https://doi.org/10.1021/acs.jpclett.7b02423} {\bibfield  {journal} {\bibinfo
   {journal} {J. Phys. Chem. Lett.}\ }\textbf {\bibinfo {volume} {8}},\
  \bibinfo {pages} {4720} (\bibinfo {year} {2017})}\BibitemShut {NoStop}%
\bibitem [{\citenamefont {{Gold-Parker}}\ \emph {et~al.}(2018)\citenamefont
  {{Gold-Parker}}, \citenamefont {Gehring}, \citenamefont {Skelton},
  \citenamefont {Smith}, \citenamefont {Parshall}, \citenamefont {Frost},
  \citenamefont {Karunadasa}, \citenamefont {Walsh},\ and\ \citenamefont
  {Toney}}]{gold-parker_AcousticPhononLifetimes_2018a}%
  \BibitemOpen
  \bibfield  {author} {\bibinfo {author} {\bibfnamefont {A.}~\bibnamefont
  {{Gold-Parker}}}, \bibinfo {author} {\bibfnamefont {P.~M.}\ \bibnamefont
  {Gehring}}, \bibinfo {author} {\bibfnamefont {J.~M.}\ \bibnamefont
  {Skelton}}, \bibinfo {author} {\bibfnamefont {I.~C.}\ \bibnamefont {Smith}},
  \bibinfo {author} {\bibfnamefont {D.}~\bibnamefont {Parshall}}, \bibinfo
  {author} {\bibfnamefont {J.~M.}\ \bibnamefont {Frost}}, \bibinfo {author}
  {\bibfnamefont {H.~I.}\ \bibnamefont {Karunadasa}}, \bibinfo {author}
  {\bibfnamefont {A.}~\bibnamefont {Walsh}},\ and\ \bibinfo {author}
  {\bibfnamefont {M.~F.}\ \bibnamefont {Toney}},\ }\href
  {https://doi.org/10.1073/pnas.1812227115} {\bibfield  {journal} {\bibinfo
  {journal} {Proc. Natl. Acad. Sci. U.S.A.}\ }\textbf {\bibinfo {volume}
  {115}},\ \bibinfo {pages} {11905} (\bibinfo {year} {2018})}\BibitemShut
  {NoStop}%
\bibitem [{\citenamefont
  {Klarbring}(2019)}]{klarbring_LowenergyPathsOctahedral_2019}%
  \BibitemOpen
  \bibfield  {author} {\bibinfo {author} {\bibfnamefont {J.}~\bibnamefont
  {Klarbring}},\ }\href {https://doi.org/10.1103/PhysRevB.99.104105} {\bibfield
   {journal} {\bibinfo  {journal} {Phys. Rev. B}\ }\textbf {\bibinfo {volume}
  {99}},\ \bibinfo {pages} {104105} (\bibinfo {year} {2019})}\BibitemShut
  {NoStop}%
\bibitem [{\citenamefont {Liu}\ \emph {et~al.}(2019)\citenamefont {Liu},
  \citenamefont {Phillips}, \citenamefont {Keen},\ and\ \citenamefont
  {Dove}}]{liu_ThermalDisorderBond_2019}%
  \BibitemOpen
  \bibfield  {author} {\bibinfo {author} {\bibfnamefont {J.}~\bibnamefont
  {Liu}}, \bibinfo {author} {\bibfnamefont {A.~E.}\ \bibnamefont {Phillips}},
  \bibinfo {author} {\bibfnamefont {D.~A.}\ \bibnamefont {Keen}},\ and\
  \bibinfo {author} {\bibfnamefont {M.~T.}\ \bibnamefont {Dove}},\ }\href
  {https://doi.org/10.1021/acs.jpcc.9b02936} {\bibfield  {journal} {\bibinfo
  {journal} {J. Phys. Chem. C}\ }\textbf {\bibinfo {volume} {123}},\ \bibinfo
  {pages} {14934} (\bibinfo {year} {2019})}\BibitemShut {NoStop}%
\bibitem [{\citenamefont {Ferreira}\ \emph {et~al.}(2020)\citenamefont
  {Ferreira}, \citenamefont {Paofai}, \citenamefont {L{\'e}toublon},
  \citenamefont {Ollivier}, \citenamefont {Raymond}, \citenamefont {Hehlen},
  \citenamefont {Ruffl{\'e}}, \citenamefont {Cordier}, \citenamefont {Katan},
  \citenamefont {Even},\ and\ \citenamefont
  {Bourges}}]{ferreira_DirectEvidenceWeakly_2020}%
  \BibitemOpen
  \bibfield  {author} {\bibinfo {author} {\bibfnamefont {A.~C.}\ \bibnamefont
  {Ferreira}}, \bibinfo {author} {\bibfnamefont {S.}~\bibnamefont {Paofai}},
  \bibinfo {author} {\bibfnamefont {A.}~\bibnamefont {L{\'e}toublon}}, \bibinfo
  {author} {\bibfnamefont {J.}~\bibnamefont {Ollivier}}, \bibinfo {author}
  {\bibfnamefont {S.}~\bibnamefont {Raymond}}, \bibinfo {author} {\bibfnamefont
  {B.}~\bibnamefont {Hehlen}}, \bibinfo {author} {\bibfnamefont
  {B.}~\bibnamefont {Ruffl{\'e}}}, \bibinfo {author} {\bibfnamefont
  {S.}~\bibnamefont {Cordier}}, \bibinfo {author} {\bibfnamefont
  {C.}~\bibnamefont {Katan}}, \bibinfo {author} {\bibfnamefont
  {J.}~\bibnamefont {Even}},\ and\ \bibinfo {author} {\bibfnamefont
  {P.}~\bibnamefont {Bourges}},\ }\href
  {https://doi.org/10.1038/s42005-020-0313-7} {\bibfield  {journal} {\bibinfo
  {journal} {Commun. Phys.}\ }\textbf {\bibinfo {volume} {3}},\ \bibinfo
  {pages} {48} (\bibinfo {year} {2020})}\BibitemShut {NoStop}%
\bibitem [{\citenamefont {{Lanigan-Atkins}}\ \emph {et~al.}(2021)\citenamefont
  {{Lanigan-Atkins}}, \citenamefont {He}, \citenamefont {Krogstad},
  \citenamefont {Pajerowski}, \citenamefont {Abernathy}, \citenamefont {Xu},
  \citenamefont {Xu}, \citenamefont {Chung}, \citenamefont {Kanatzidis},
  \citenamefont {Rosenkranz}, \citenamefont {Osborn},\ and\ \citenamefont
  {Delaire}}]{lanigan-atkins_TwodimensionalOverdampedFluctuations_2021}%
  \BibitemOpen
  \bibfield  {author} {\bibinfo {author} {\bibfnamefont {T.}~\bibnamefont
  {{Lanigan-Atkins}}}, \bibinfo {author} {\bibfnamefont {X.}~\bibnamefont
  {He}}, \bibinfo {author} {\bibfnamefont {M.~J.}\ \bibnamefont {Krogstad}},
  \bibinfo {author} {\bibfnamefont {D.~M.}\ \bibnamefont {Pajerowski}},
  \bibinfo {author} {\bibfnamefont {D.~L.}\ \bibnamefont {Abernathy}}, \bibinfo
  {author} {\bibfnamefont {G.~N. M.~N.}\ \bibnamefont {Xu}}, \bibinfo {author}
  {\bibfnamefont {Z.}~\bibnamefont {Xu}}, \bibinfo {author} {\bibfnamefont
  {D.-Y.}\ \bibnamefont {Chung}}, \bibinfo {author} {\bibfnamefont {M.~G.}\
  \bibnamefont {Kanatzidis}}, \bibinfo {author} {\bibfnamefont
  {S.}~\bibnamefont {Rosenkranz}}, \bibinfo {author} {\bibfnamefont
  {R.}~\bibnamefont {Osborn}},\ and\ \bibinfo {author} {\bibfnamefont
  {O.}~\bibnamefont {Delaire}},\ }\href
  {https://doi.org/10.1038/s41563-021-00947-y} {\bibfield  {journal} {\bibinfo
  {journal} {Nat. Mater.}\ }\textbf {\bibinfo {volume} {20}},\ \bibinfo {pages}
  {977} (\bibinfo {year} {2021})}\BibitemShut {NoStop}%
\bibitem [{\citenamefont {Debnath}\ \emph {et~al.}(2021)\citenamefont
  {Debnath}, \citenamefont {Sarker}, \citenamefont {Huang}, \citenamefont
  {Han}, \citenamefont {Dey}, \citenamefont {Polavarapu}, \citenamefont
  {Levchenko},\ and\ \citenamefont
  {Feldmann}}]{debnath_CoherentVibrationalDynamics_2021}%
  \BibitemOpen
  \bibfield  {author} {\bibinfo {author} {\bibfnamefont {T.}~\bibnamefont
  {Debnath}}, \bibinfo {author} {\bibfnamefont {D.}~\bibnamefont {Sarker}},
  \bibinfo {author} {\bibfnamefont {H.}~\bibnamefont {Huang}}, \bibinfo
  {author} {\bibfnamefont {Z.-K.}\ \bibnamefont {Han}}, \bibinfo {author}
  {\bibfnamefont {A.}~\bibnamefont {Dey}}, \bibinfo {author} {\bibfnamefont
  {L.}~\bibnamefont {Polavarapu}}, \bibinfo {author} {\bibfnamefont {S.~V.}\
  \bibnamefont {Levchenko}},\ and\ \bibinfo {author} {\bibfnamefont
  {J.}~\bibnamefont {Feldmann}},\ }\href
  {https://doi.org/10.1038/s41467-021-22934-2} {\bibfield  {journal} {\bibinfo
  {journal} {Nat. Commun.}\ }\textbf {\bibinfo {volume} {12}},\ \bibinfo
  {pages} {2629} (\bibinfo {year} {2021})}\BibitemShut {NoStop}%
\bibitem [{\citenamefont {Cannelli}\ \emph {et~al.}(2022)\citenamefont
  {Cannelli}, \citenamefont {Wiktor}, \citenamefont {Colonna}, \citenamefont
  {Leroy}, \citenamefont {Puppin}, \citenamefont {Bacellar}, \citenamefont
  {Sadykov}, \citenamefont {Krieg}, \citenamefont {Smolentsev}, \citenamefont
  {Kovalenko}, \citenamefont {Pasquarello}, \citenamefont {Chergui},\ and\
  \citenamefont {Mancini}}]{cannelli_AtomicLevelDescriptionThermal_2022}%
  \BibitemOpen
  \bibfield  {author} {\bibinfo {author} {\bibfnamefont {O.}~\bibnamefont
  {Cannelli}}, \bibinfo {author} {\bibfnamefont {J.}~\bibnamefont {Wiktor}},
  \bibinfo {author} {\bibfnamefont {N.}~\bibnamefont {Colonna}}, \bibinfo
  {author} {\bibfnamefont {L.}~\bibnamefont {Leroy}}, \bibinfo {author}
  {\bibfnamefont {M.}~\bibnamefont {Puppin}}, \bibinfo {author} {\bibfnamefont
  {C.}~\bibnamefont {Bacellar}}, \bibinfo {author} {\bibfnamefont
  {I.}~\bibnamefont {Sadykov}}, \bibinfo {author} {\bibfnamefont
  {F.}~\bibnamefont {Krieg}}, \bibinfo {author} {\bibfnamefont
  {G.}~\bibnamefont {Smolentsev}}, \bibinfo {author} {\bibfnamefont {M.~V.}\
  \bibnamefont {Kovalenko}}, \bibinfo {author} {\bibfnamefont {A.}~\bibnamefont
  {Pasquarello}}, \bibinfo {author} {\bibfnamefont {M.}~\bibnamefont
  {Chergui}},\ and\ \bibinfo {author} {\bibfnamefont {G.~F.}\ \bibnamefont
  {Mancini}},\ }\href {https://doi.org/10.1021/acs.jpclett.2c00281} {\bibfield
  {journal} {\bibinfo  {journal} {J. Phys. Chem. Lett.}\ }\textbf {\bibinfo
  {volume} {13}},\ \bibinfo {pages} {3382} (\bibinfo {year}
  {2022})}\BibitemShut {NoStop}%
\bibitem [{\citenamefont {Zhu}\ \emph {et~al.}(2022)\citenamefont {Zhu},
  \citenamefont {{Caicedo-D{\'a}vila}}, \citenamefont {Gehrmann},\ and\
  \citenamefont {Egger}}]{zhu_ProbingDisorderCubic_2022}%
  \BibitemOpen
  \bibfield  {author} {\bibinfo {author} {\bibfnamefont {X.}~\bibnamefont
  {Zhu}}, \bibinfo {author} {\bibfnamefont {S.}~\bibnamefont
  {{Caicedo-D{\'a}vila}}}, \bibinfo {author} {\bibfnamefont {C.}~\bibnamefont
  {Gehrmann}},\ and\ \bibinfo {author} {\bibfnamefont {D.~A.}\ \bibnamefont
  {Egger}},\ }\href {https://doi.org/10.1021/acsami.1c23099} {\bibfield
  {journal} {\bibinfo  {journal} {ACS Appl. Mater. Interfaces}\ }\textbf
  {\bibinfo {volume} {14}},\ \bibinfo {pages} {22973} (\bibinfo {year}
  {2022})}\BibitemShut {NoStop}%
\bibitem [{\citenamefont {Fransson}\ \emph {et~al.}(2023)\citenamefont
  {Fransson}, \citenamefont {Rosander}, \citenamefont {Eriksson}, \citenamefont
  {Rahm}, \citenamefont {Tadano},\ and\ \citenamefont
  {Erhart}}]{fransson_LimitsPhononQuasiparticle_2023}%
  \BibitemOpen
  \bibfield  {author} {\bibinfo {author} {\bibfnamefont {E.}~\bibnamefont
  {Fransson}}, \bibinfo {author} {\bibfnamefont {P.}~\bibnamefont {Rosander}},
  \bibinfo {author} {\bibfnamefont {F.}~\bibnamefont {Eriksson}}, \bibinfo
  {author} {\bibfnamefont {J.~M.}\ \bibnamefont {Rahm}}, \bibinfo {author}
  {\bibfnamefont {T.}~\bibnamefont {Tadano}},\ and\ \bibinfo {author}
  {\bibfnamefont {P.}~\bibnamefont {Erhart}},\ }\href
  {https://doi.org/10.1038/s42005-023-01297-8} {\bibfield  {journal} {\bibinfo
  {journal} {Commun. Phys.}\ }\textbf {\bibinfo {volume} {6}},\ \bibinfo
  {pages} {173} (\bibinfo {year} {2023})}\BibitemShut {NoStop}%
\bibitem [{\citenamefont {Weadock}\ \emph {et~al.}(2023)\citenamefont
  {Weadock}, \citenamefont {Sterling}, \citenamefont {Vigil}, \citenamefont
  {{Gold-Parker}}, \citenamefont {Smith}, \citenamefont {Ahammed},
  \citenamefont {Krogstad}, \citenamefont {Ye}, \citenamefont {Voneshen},
  \citenamefont {Gehring}, \citenamefont {Rappe}, \citenamefont
  {Steinr{\"u}ck}, \citenamefont {Ertekin}, \citenamefont {Karunadasa},
  \citenamefont {Reznik},\ and\ \citenamefont
  {Toney}}]{weadock_NatureDynamicLocal_2023a}%
  \BibitemOpen
  \bibfield  {author} {\bibinfo {author} {\bibfnamefont {N.~J.}\ \bibnamefont
  {Weadock}}, \bibinfo {author} {\bibfnamefont {T.~C.}\ \bibnamefont
  {Sterling}}, \bibinfo {author} {\bibfnamefont {J.~A.}\ \bibnamefont {Vigil}},
  \bibinfo {author} {\bibfnamefont {A.}~\bibnamefont {{Gold-Parker}}}, \bibinfo
  {author} {\bibfnamefont {I.~C.}\ \bibnamefont {Smith}}, \bibinfo {author}
  {\bibfnamefont {B.}~\bibnamefont {Ahammed}}, \bibinfo {author} {\bibfnamefont
  {M.~J.}\ \bibnamefont {Krogstad}}, \bibinfo {author} {\bibfnamefont
  {F.}~\bibnamefont {Ye}}, \bibinfo {author} {\bibfnamefont {D.}~\bibnamefont
  {Voneshen}}, \bibinfo {author} {\bibfnamefont {P.~M.}\ \bibnamefont
  {Gehring}}, \bibinfo {author} {\bibfnamefont {A.~M.}\ \bibnamefont {Rappe}},
  \bibinfo {author} {\bibfnamefont {H.-G.}\ \bibnamefont {Steinr{\"u}ck}},
  \bibinfo {author} {\bibfnamefont {E.}~\bibnamefont {Ertekin}}, \bibinfo
  {author} {\bibfnamefont {H.~I.}\ \bibnamefont {Karunadasa}}, \bibinfo
  {author} {\bibfnamefont {D.}~\bibnamefont {Reznik}},\ and\ \bibinfo {author}
  {\bibfnamefont {M.~F.}\ \bibnamefont {Toney}},\ }\href
  {https://doi.org/10.1016/j.joule.2023.03.017} {\bibfield  {journal} {\bibinfo
   {journal} {Joule}\ }\textbf {\bibinfo {volume} {7}},\ \bibinfo {pages}
  {1051} (\bibinfo {year} {2023})}\BibitemShut {NoStop}%
\bibitem [{\citenamefont {Irvine}\ \emph {et~al.}(2021)\citenamefont {Irvine},
  \citenamefont {Walker},\ and\ \citenamefont
  {Wolf}}]{irvine_QuantifyingPolaronicEffects_2021}%
  \BibitemOpen
  \bibfield  {author} {\bibinfo {author} {\bibfnamefont {L.~A.~D.}\
  \bibnamefont {Irvine}}, \bibinfo {author} {\bibfnamefont {A.~B.}\
  \bibnamefont {Walker}},\ and\ \bibinfo {author} {\bibfnamefont {M.~J.}\
  \bibnamefont {Wolf}},\ }\href {https://doi.org/10.1103/PhysRevB.103.L220305}
  {\bibfield  {journal} {\bibinfo  {journal} {Phys. Rev. B}\ }\textbf {\bibinfo
  {volume} {103}},\ \bibinfo {pages} {L220305} (\bibinfo {year}
  {2021})}\BibitemShut {NoStop}%
\bibitem [{\citenamefont {Zhang}\ \emph {et~al.}(2022)\citenamefont {Zhang},
  \citenamefont {Shen}, \citenamefont {Zhang}, \citenamefont {Li},\ and\
  \citenamefont {Liu}}]{zhang_EffectQuarticAnharmonicity_2022}%
  \BibitemOpen
  \bibfield  {author} {\bibinfo {author} {\bibfnamefont {K.-C.}\ \bibnamefont
  {Zhang}}, \bibinfo {author} {\bibfnamefont {C.}~\bibnamefont {Shen}},
  \bibinfo {author} {\bibfnamefont {H.-B.}\ \bibnamefont {Zhang}}, \bibinfo
  {author} {\bibfnamefont {Y.-F.}\ \bibnamefont {Li}},\ and\ \bibinfo {author}
  {\bibfnamefont {Y.}~\bibnamefont {Liu}},\ }\href
  {https://doi.org/10.1103/PhysRevB.106.235202} {\bibfield  {journal} {\bibinfo
   {journal} {Phys. Rev. B}\ }\textbf {\bibinfo {volume} {106}},\ \bibinfo
  {pages} {235202} (\bibinfo {year} {2022})}\BibitemShut {NoStop}%
\bibitem [{\citenamefont {Zacharias}\ \emph {et~al.}(2023)\citenamefont
  {Zacharias}, \citenamefont {Volonakis}, \citenamefont {Giustino},\ and\
  \citenamefont {Even}}]{zacharias_AnharmonicElectronphononCoupling_2023}%
  \BibitemOpen
  \bibfield  {author} {\bibinfo {author} {\bibfnamefont {M.}~\bibnamefont
  {Zacharias}}, \bibinfo {author} {\bibfnamefont {G.}~\bibnamefont
  {Volonakis}}, \bibinfo {author} {\bibfnamefont {F.}~\bibnamefont
  {Giustino}},\ and\ \bibinfo {author} {\bibfnamefont {J.}~\bibnamefont
  {Even}},\ }\Eprint {https://arxiv.org/abs/2302.09625} {arxiv:2302.09625
  [cond-mat]}  (\bibinfo {year} {2023})\BibitemShut {NoStop}%
\bibitem [{\citenamefont {Seidl}\ \emph {et~al.}(2023)\citenamefont {Seidl},
  \citenamefont {Zhu}, \citenamefont {Reuveni}, \citenamefont {Aharon},
  \citenamefont {Gehrmann}, \citenamefont {{Caicedo-D{\'a}vila}}, \citenamefont
  {Yaffe},\ and\ \citenamefont
  {Egger}}]{seidl_AnharmonicFluctuationsGovern_2023}%
  \BibitemOpen
  \bibfield  {author} {\bibinfo {author} {\bibfnamefont {S.~A.}\ \bibnamefont
  {Seidl}}, \bibinfo {author} {\bibfnamefont {X.}~\bibnamefont {Zhu}}, \bibinfo
  {author} {\bibfnamefont {G.}~\bibnamefont {Reuveni}}, \bibinfo {author}
  {\bibfnamefont {S.}~\bibnamefont {Aharon}}, \bibinfo {author} {\bibfnamefont
  {C.}~\bibnamefont {Gehrmann}}, \bibinfo {author} {\bibfnamefont
  {S.}~\bibnamefont {{Caicedo-D{\'a}vila}}}, \bibinfo {author} {\bibfnamefont
  {O.}~\bibnamefont {Yaffe}},\ and\ \bibinfo {author} {\bibfnamefont {D.~A.}\
  \bibnamefont {Egger}},\ }\Eprint {https://arxiv.org/abs/2303.01603}
  {arxiv:2303.01603 [cond-mat]}  (\bibinfo {year} {2023})\BibitemShut {NoStop}%
\bibitem [{\citenamefont {Karakus}\ \emph {et~al.}(2015)\citenamefont
  {Karakus}, \citenamefont {Jensen}, \citenamefont {D'Angelo}, \citenamefont
  {Turchinovich}, \citenamefont {Bonn},\ and\ \citenamefont
  {C{\'a}novas}}]{karakus_PhononElectronScattering_2015a}%
  \BibitemOpen
  \bibfield  {author} {\bibinfo {author} {\bibfnamefont {M.}~\bibnamefont
  {Karakus}}, \bibinfo {author} {\bibfnamefont {S.~A.}\ \bibnamefont {Jensen}},
  \bibinfo {author} {\bibfnamefont {F.}~\bibnamefont {D'Angelo}}, \bibinfo
  {author} {\bibfnamefont {D.}~\bibnamefont {Turchinovich}}, \bibinfo {author}
  {\bibfnamefont {M.}~\bibnamefont {Bonn}},\ and\ \bibinfo {author}
  {\bibfnamefont {E.}~\bibnamefont {C{\'a}novas}},\ }\href
  {https://doi.org/10.1021/acs.jpclett.5b02485} {\bibfield  {journal} {\bibinfo
   {journal} {J. Phys. Chem. Lett.}\ }\textbf {\bibinfo {volume} {6}},\
  \bibinfo {pages} {4991} (\bibinfo {year} {2015})}\BibitemShut {NoStop}%
\bibitem [{\citenamefont {Hill}\ \emph {et~al.}(2018)\citenamefont {Hill},
  \citenamefont {Kennedy}, \citenamefont {Massaro},\ and\ \citenamefont
  {Grumstrup}}]{hill_PerovskiteCarrierTransport_2018}%
  \BibitemOpen
  \bibfield  {author} {\bibinfo {author} {\bibfnamefont {A.~H.}\ \bibnamefont
  {Hill}}, \bibinfo {author} {\bibfnamefont {C.~L.}\ \bibnamefont {Kennedy}},
  \bibinfo {author} {\bibfnamefont {E.~S.}\ \bibnamefont {Massaro}},\ and\
  \bibinfo {author} {\bibfnamefont {E.~M.}\ \bibnamefont {Grumstrup}},\ }\href
  {https://doi.org/10.1021/acs.jpclett.8b00652} {\bibfield  {journal} {\bibinfo
   {journal} {J. Phys. Chem. Lett.}\ }\textbf {\bibinfo {volume} {9}},\
  \bibinfo {pages} {2808} (\bibinfo {year} {2018})}\BibitemShut {NoStop}%
\bibitem [{\citenamefont {Lacroix}\ \emph {et~al.}(2020)\citenamefont
  {Lacroix}, \citenamefont {{de Laissardi{\`e}re}}, \citenamefont
  {Qu{\'e}merais}, \citenamefont {Julien},\ and\ \citenamefont
  {Mayou}}]{lacroix_ModelingElectronicMobilities_2020a}%
  \BibitemOpen
  \bibfield  {author} {\bibinfo {author} {\bibfnamefont {A.}~\bibnamefont
  {Lacroix}}, \bibinfo {author} {\bibfnamefont {G.~T.}\ \bibnamefont {{de
  Laissardi{\`e}re}}}, \bibinfo {author} {\bibfnamefont {P.}~\bibnamefont
  {Qu{\'e}merais}}, \bibinfo {author} {\bibfnamefont {J.-P.}\ \bibnamefont
  {Julien}},\ and\ \bibinfo {author} {\bibfnamefont {D.}~\bibnamefont
  {Mayou}},\ }\href {https://doi.org/10.1103/PhysRevLett.124.196601} {\bibfield
   {journal} {\bibinfo  {journal} {Phys. Rev. Lett.}\ }\textbf {\bibinfo
  {volume} {124}},\ \bibinfo {pages} {196601} (\bibinfo {year}
  {2020})}\BibitemShut {NoStop}%
\bibitem [{\citenamefont {Ioffe}\ and\ \citenamefont
  {Regel}(1960)}]{ioffe_NoncrystallineAmorphousLiquid_1960}%
  \BibitemOpen
  \bibfield  {author} {\bibinfo {author} {\bibfnamefont {A.~F.}\ \bibnamefont
  {Ioffe}}\ and\ \bibinfo {author} {\bibfnamefont {A.~R.}\ \bibnamefont
  {Regel}},\ }\href@noop {} {\bibfield  {journal} {\bibinfo  {journal} {Prog.
  Semicond}\ }\textbf {\bibinfo {volume} {4}},\ \bibinfo {pages} {237}
  (\bibinfo {year} {1960})}\BibitemShut {NoStop}%
\bibitem [{\citenamefont
  {Mott}(1972)}]{mott_ConductionNoncrystallineSystems_1972}%
  \BibitemOpen
  \bibfield  {author} {\bibinfo {author} {\bibfnamefont {N.~F.}\ \bibnamefont
  {Mott}},\ }\href {https://doi.org/10.1080/14786437208226973} {\bibfield
  {journal} {\bibinfo  {journal} {Philos. Mag.}\ }\textbf {\bibinfo {volume}
  {26}},\ \bibinfo {pages} {1015} (\bibinfo {year} {1972})}\BibitemShut
  {NoStop}%
\bibitem [{\citenamefont {Biewald}\ \emph {et~al.}(2019)\citenamefont
  {Biewald}, \citenamefont {Giesbrecht}, \citenamefont {Bein}, \citenamefont
  {Docampo}, \citenamefont {Hartschuh},\ and\ \citenamefont
  {Ciesielski}}]{biewald_TemperatureDependentAmbipolarCharge_2019a}%
  \BibitemOpen
  \bibfield  {author} {\bibinfo {author} {\bibfnamefont {A.}~\bibnamefont
  {Biewald}}, \bibinfo {author} {\bibfnamefont {N.}~\bibnamefont {Giesbrecht}},
  \bibinfo {author} {\bibfnamefont {T.}~\bibnamefont {Bein}}, \bibinfo {author}
  {\bibfnamefont {P.}~\bibnamefont {Docampo}}, \bibinfo {author} {\bibfnamefont
  {A.}~\bibnamefont {Hartschuh}},\ and\ \bibinfo {author} {\bibfnamefont
  {R.}~\bibnamefont {Ciesielski}},\ }\href
  {https://doi.org/10.1021/acsami.9b04592} {\bibfield  {journal} {\bibinfo
  {journal} {ACS Appl. Mater. Interfaces}\ }\textbf {\bibinfo {volume} {11}},\
  \bibinfo {pages} {20838} (\bibinfo {year} {2019})}\BibitemShut {NoStop}%
\bibitem [{\citenamefont {Mayers}\ \emph {et~al.}(2018)\citenamefont {Mayers},
  \citenamefont {Tan}, \citenamefont {Egger}, \citenamefont {Rappe},\ and\
  \citenamefont {Reichman}}]{mayers_HowLatticeCharge_2018}%
  \BibitemOpen
  \bibfield  {author} {\bibinfo {author} {\bibfnamefont {M.~Z.}\ \bibnamefont
  {Mayers}}, \bibinfo {author} {\bibfnamefont {L.~Z.}\ \bibnamefont {Tan}},
  \bibinfo {author} {\bibfnamefont {D.~A.}\ \bibnamefont {Egger}}, \bibinfo
  {author} {\bibfnamefont {A.~M.}\ \bibnamefont {Rappe}},\ and\ \bibinfo
  {author} {\bibfnamefont {D.~R.}\ \bibnamefont {Reichman}},\ }\href
  {https://doi.org/10.1021/acs.nanolett.8b04276} {\bibfield  {journal}
  {\bibinfo  {journal} {Nano Lett.}\ }\textbf {\bibinfo {volume} {18}},\
  \bibinfo {pages} {8041} (\bibinfo {year} {2018})}\BibitemShut {NoStop}%
\bibitem [{\citenamefont {Mishchenko}\ \emph {et~al.}(2019)\citenamefont
  {Mishchenko}, \citenamefont {Pollet}, \citenamefont {Prokof'ev},
  \citenamefont {Kumar}, \citenamefont {Maslov},\ and\ \citenamefont
  {Nagaosa}}]{mishchenko_PolaronMobilityQuasiparticles_2019}%
  \BibitemOpen
  \bibfield  {author} {\bibinfo {author} {\bibfnamefont {A.~S.}\ \bibnamefont
  {Mishchenko}}, \bibinfo {author} {\bibfnamefont {L.}~\bibnamefont {Pollet}},
  \bibinfo {author} {\bibfnamefont {N.~V.}\ \bibnamefont {Prokof'ev}}, \bibinfo
  {author} {\bibfnamefont {A.}~\bibnamefont {Kumar}}, \bibinfo {author}
  {\bibfnamefont {D.~L.}\ \bibnamefont {Maslov}},\ and\ \bibinfo {author}
  {\bibfnamefont {N.}~\bibnamefont {Nagaosa}},\ }\href
  {https://doi.org/10.1103/PhysRevLett.123.076601} {\bibfield  {journal}
  {\bibinfo  {journal} {Phys. Rev. Lett.}\ }\textbf {\bibinfo {volume} {123}},\
  \bibinfo {pages} {076601} (\bibinfo {year} {2019})}\BibitemShut {NoStop}%
\bibitem [{\citenamefont {Wright}\ \emph {et~al.}(2016)\citenamefont {Wright},
  \citenamefont {Verdi}, \citenamefont {Milot}, \citenamefont {Eperon},
  \citenamefont {{P{\'e}rez-Osorio}}, \citenamefont {Snaith}, \citenamefont
  {Giustino}, \citenamefont {Johnston},\ and\ \citenamefont
  {Herz}}]{wright_ElectronPhononCoupling_2016}%
  \BibitemOpen
  \bibfield  {author} {\bibinfo {author} {\bibfnamefont {A.~D.}\ \bibnamefont
  {Wright}}, \bibinfo {author} {\bibfnamefont {C.}~\bibnamefont {Verdi}},
  \bibinfo {author} {\bibfnamefont {R.~L.}\ \bibnamefont {Milot}}, \bibinfo
  {author} {\bibfnamefont {G.~E.}\ \bibnamefont {Eperon}}, \bibinfo {author}
  {\bibfnamefont {M.~A.}\ \bibnamefont {{P{\'e}rez-Osorio}}}, \bibinfo {author}
  {\bibfnamefont {H.~J.}\ \bibnamefont {Snaith}}, \bibinfo {author}
  {\bibfnamefont {F.}~\bibnamefont {Giustino}}, \bibinfo {author}
  {\bibfnamefont {M.~B.}\ \bibnamefont {Johnston}},\ and\ \bibinfo {author}
  {\bibfnamefont {L.~M.}\ \bibnamefont {Herz}},\ }\href
  {https://doi.org/10.1038/ncomms11755} {\bibfield  {journal} {\bibinfo
  {journal} {Nat. Commun.}\ }\textbf {\bibinfo {volume} {7}},\ \bibinfo {pages}
  {11755} (\bibinfo {year} {2016})}\BibitemShut {NoStop}%
\bibitem [{\citenamefont {Martin}\ and\ \citenamefont
  {Frost}(2023)}]{martin_MultiplePhononModes_2023}%
  \BibitemOpen
  \bibfield  {author} {\bibinfo {author} {\bibfnamefont {B.~A.~A.}\
  \bibnamefont {Martin}}\ and\ \bibinfo {author} {\bibfnamefont {J.~M.}\
  \bibnamefont {Frost}},\ }\href {https://doi.org/10.1103/PhysRevB.107.115203}
  {\bibfield  {journal} {\bibinfo  {journal} {Phys. Rev. B}\ }\textbf {\bibinfo
  {volume} {107}},\ \bibinfo {pages} {115203} (\bibinfo {year}
  {2023})}\BibitemShut {NoStop}%
\bibitem [{\citenamefont {Abramovitch}\ \emph {et~al.}(2021)\citenamefont
  {Abramovitch}, \citenamefont {Saidi},\ and\ \citenamefont
  {Tan}}]{abramovitch_ThermalFluctuationsCarrier_2021}%
  \BibitemOpen
  \bibfield  {author} {\bibinfo {author} {\bibfnamefont {D.~J.}\ \bibnamefont
  {Abramovitch}}, \bibinfo {author} {\bibfnamefont {W.~A.}\ \bibnamefont
  {Saidi}},\ and\ \bibinfo {author} {\bibfnamefont {L.~Z.}\ \bibnamefont
  {Tan}},\ }\href {https://doi.org/10.1103/PhysRevMaterials.5.085404}
  {\bibfield  {journal} {\bibinfo  {journal} {Phys. Rev. Mater.}\ }\textbf
  {\bibinfo {volume} {5}},\ \bibinfo {pages} {085404} (\bibinfo {year}
  {2021})}\BibitemShut {NoStop}%
\bibitem [{\citenamefont {Troisi}\ and\ \citenamefont
  {Orlandi}(2006)}]{troisi_ChargeTransportRegimeCrystalline_2006}%
  \BibitemOpen
  \bibfield  {author} {\bibinfo {author} {\bibfnamefont {A.}~\bibnamefont
  {Troisi}}\ and\ \bibinfo {author} {\bibfnamefont {G.}~\bibnamefont
  {Orlandi}},\ }\href {https://doi.org/10.1103/PhysRevLett.96.086601}
  {\bibfield  {journal} {\bibinfo  {journal} {Phys. Rev. Lett.}\ }\textbf
  {\bibinfo {volume} {96}},\ \bibinfo {pages} {086601} (\bibinfo {year}
  {2006})}\BibitemShut {NoStop}%
\bibitem [{\citenamefont {Troisi}(2011)}]{troisi_ChargeTransportHigh_2011a}%
  \BibitemOpen
  \bibfield  {author} {\bibinfo {author} {\bibfnamefont {A.}~\bibnamefont
  {Troisi}},\ }\href {https://doi.org/10.1039/c0cs00198h} {\bibfield  {journal}
  {\bibinfo  {journal} {Chem. Soc. Rev.}\ }\textbf {\bibinfo {volume} {40}},\
  \bibinfo {pages} {2347} (\bibinfo {year} {2011})}\BibitemShut {NoStop}%
\bibitem [{\citenamefont {Wang}\ and\ \citenamefont
  {Beljonne}(2013)}]{wang_ChargeTransportOrganic_2013}%
  \BibitemOpen
  \bibfield  {author} {\bibinfo {author} {\bibfnamefont {L.}~\bibnamefont
  {Wang}}\ and\ \bibinfo {author} {\bibfnamefont {D.}~\bibnamefont
  {Beljonne}},\ }\href {https://doi.org/10.1063/1.4817856} {\bibfield
  {journal} {\bibinfo  {journal} {J. Chem. Phys.}\ }\textbf {\bibinfo {volume}
  {139}},\ \bibinfo {pages} {064316} (\bibinfo {year} {2013})}\BibitemShut
  {NoStop}%
\bibitem [{\citenamefont {Fratini}\ \emph {et~al.}(2016)\citenamefont
  {Fratini}, \citenamefont {Mayou},\ and\ \citenamefont
  {Ciuchi}}]{fratini_TransientLocalizationScenario_2016}%
  \BibitemOpen
  \bibfield  {author} {\bibinfo {author} {\bibfnamefont {S.}~\bibnamefont
  {Fratini}}, \bibinfo {author} {\bibfnamefont {D.}~\bibnamefont {Mayou}},\
  and\ \bibinfo {author} {\bibfnamefont {S.}~\bibnamefont {Ciuchi}},\ }\href
  {https://doi.org/10.1002/adfm.201502386} {\bibfield  {journal} {\bibinfo
  {journal} {Adv. Funct. Mater.}\ }\textbf {\bibinfo {volume} {26}},\ \bibinfo
  {pages} {2292} (\bibinfo {year} {2016})}\BibitemShut {NoStop}%
\bibitem [{\citenamefont {Fratini}\ \emph {et~al.}(2017)\citenamefont
  {Fratini}, \citenamefont {Ciuchi}, \citenamefont {Mayou}, \citenamefont {{de
  Laissardi{\`e}re}},\ and\ \citenamefont
  {Troisi}}]{fratini_MapHighmobilityMolecular_2017}%
  \BibitemOpen
  \bibfield  {author} {\bibinfo {author} {\bibfnamefont {S.}~\bibnamefont
  {Fratini}}, \bibinfo {author} {\bibfnamefont {S.}~\bibnamefont {Ciuchi}},
  \bibinfo {author} {\bibfnamefont {D.}~\bibnamefont {Mayou}}, \bibinfo
  {author} {\bibfnamefont {G.~T.}\ \bibnamefont {{de Laissardi{\`e}re}}},\ and\
  \bibinfo {author} {\bibfnamefont {A.}~\bibnamefont {Troisi}},\ }\href
  {https://doi.org/10.1038/nmat4970} {\bibfield  {journal} {\bibinfo  {journal}
  {Nat. Mater.}\ }\textbf {\bibinfo {volume} {16}},\ \bibinfo {pages} {998}
  (\bibinfo {year} {2017})}\BibitemShut {NoStop}%
\bibitem [{\citenamefont {Asher}\ \emph {et~al.}(2020)\citenamefont {Asher},
  \citenamefont {Angerer}, \citenamefont {Korobko}, \citenamefont
  {Diskin-Posner}, \citenamefont {Egger},\ and\ \citenamefont
  {Yaffe}}]{asher_AnharmonicLatticeVibrations_2020a}%
  \BibitemOpen
  \bibfield  {author} {\bibinfo {author} {\bibfnamefont {M.}~\bibnamefont
  {Asher}}, \bibinfo {author} {\bibfnamefont {D.}~\bibnamefont {Angerer}},
  \bibinfo {author} {\bibfnamefont {R.}~\bibnamefont {Korobko}}, \bibinfo
  {author} {\bibfnamefont {Y.}~\bibnamefont {Diskin-Posner}}, \bibinfo {author}
  {\bibfnamefont {D.~A.}\ \bibnamefont {Egger}},\ and\ \bibinfo {author}
  {\bibfnamefont {O.}~\bibnamefont {Yaffe}},\ }\href
  {https://doi.org/10.1002/adma.201908028} {\bibfield  {journal} {\bibinfo
  {journal} {Adv. Mater.}\ }\textbf {\bibinfo {volume} {32}},\ \bibinfo {pages}
  {1908028} (\bibinfo {year} {2020})}\BibitemShut {NoStop}%
\bibitem [{\citenamefont {Giannini}\ \emph {et~al.}(2020)\citenamefont
  {Giannini}, \citenamefont {Ziogos}, \citenamefont {Carof}, \citenamefont
  {Ellis},\ and\ \citenamefont
  {Blumberger}}]{giannini_FlickeringPolaronsExtending_2020}%
  \BibitemOpen
  \bibfield  {author} {\bibinfo {author} {\bibfnamefont {S.}~\bibnamefont
  {Giannini}}, \bibinfo {author} {\bibfnamefont {O.~G.}\ \bibnamefont
  {Ziogos}}, \bibinfo {author} {\bibfnamefont {A.}~\bibnamefont {Carof}},
  \bibinfo {author} {\bibfnamefont {M.}~\bibnamefont {Ellis}},\ and\ \bibinfo
  {author} {\bibfnamefont {J.}~\bibnamefont {Blumberger}},\ }\href
  {https://doi.org/10.1002/adts.202000093} {\bibfield  {journal} {\bibinfo
  {journal} {Adv. Theory Simul.}\ }\textbf {\bibinfo {volume} {3}},\ \bibinfo
  {pages} {2000093} (\bibinfo {year} {2020})}\BibitemShut {NoStop}%
\bibitem [{\citenamefont {Stoeckel}\ \emph {et~al.}(2021)\citenamefont
  {Stoeckel}, \citenamefont {Olivier}, \citenamefont {Gobbi}, \citenamefont
  {Dudenko}, \citenamefont {Lemaur}, \citenamefont {Zbiri}, \citenamefont
  {Guilbert}, \citenamefont {D'Avino}, \citenamefont {Liscio}, \citenamefont
  {Migliori}, \citenamefont {Ortolani}, \citenamefont {Demitri}, \citenamefont
  {Jin}, \citenamefont {Jeong}, \citenamefont {Liscio}, \citenamefont {Nardi},
  \citenamefont {Pasquali}, \citenamefont {Razzari}, \citenamefont {Beljonne},
  \citenamefont {Samor{\`i}},\ and\ \citenamefont
  {Orgiu}}]{stoeckel_AnalysisExternalInternal_2021}%
  \BibitemOpen
  \bibfield  {author} {\bibinfo {author} {\bibfnamefont {M.-A.}\ \bibnamefont
  {Stoeckel}}, \bibinfo {author} {\bibfnamefont {Y.}~\bibnamefont {Olivier}},
  \bibinfo {author} {\bibfnamefont {M.}~\bibnamefont {Gobbi}}, \bibinfo
  {author} {\bibfnamefont {D.}~\bibnamefont {Dudenko}}, \bibinfo {author}
  {\bibfnamefont {V.}~\bibnamefont {Lemaur}}, \bibinfo {author} {\bibfnamefont
  {M.}~\bibnamefont {Zbiri}}, \bibinfo {author} {\bibfnamefont {A.~A.~Y.}\
  \bibnamefont {Guilbert}}, \bibinfo {author} {\bibfnamefont {G.}~\bibnamefont
  {D'Avino}}, \bibinfo {author} {\bibfnamefont {F.}~\bibnamefont {Liscio}},
  \bibinfo {author} {\bibfnamefont {A.}~\bibnamefont {Migliori}}, \bibinfo
  {author} {\bibfnamefont {L.}~\bibnamefont {Ortolani}}, \bibinfo {author}
  {\bibfnamefont {N.}~\bibnamefont {Demitri}}, \bibinfo {author} {\bibfnamefont
  {X.}~\bibnamefont {Jin}}, \bibinfo {author} {\bibfnamefont {Y.-G.}\
  \bibnamefont {Jeong}}, \bibinfo {author} {\bibfnamefont {A.}~\bibnamefont
  {Liscio}}, \bibinfo {author} {\bibfnamefont {M.-V.}\ \bibnamefont {Nardi}},
  \bibinfo {author} {\bibfnamefont {L.}~\bibnamefont {Pasquali}}, \bibinfo
  {author} {\bibfnamefont {L.}~\bibnamefont {Razzari}}, \bibinfo {author}
  {\bibfnamefont {D.}~\bibnamefont {Beljonne}}, \bibinfo {author}
  {\bibfnamefont {P.}~\bibnamefont {Samor{\`i}}},\ and\ \bibinfo {author}
  {\bibfnamefont {E.}~\bibnamefont {Orgiu}},\ }\href
  {https://doi.org/10.1002/adma.202007870} {\bibfield  {journal} {\bibinfo
  {journal} {Adv. Mater.}\ }\textbf {\bibinfo {volume} {33}},\ \bibinfo {pages}
  {2007870} (\bibinfo {year} {2021})}\BibitemShut {NoStop}%
\bibitem [{\citenamefont {Giannini}\ \emph {et~al.}(2023)\citenamefont
  {Giannini}, \citenamefont {Di~Virgilio}, \citenamefont {Bardini},
  \citenamefont {Hausch}, \citenamefont {Geuchies}, \citenamefont {Zheng},
  \citenamefont {Volpi}, \citenamefont {Elsner}, \citenamefont {Broch},
  \citenamefont {Geerts}, \citenamefont {Schreiber}, \citenamefont
  {Schweicher}, \citenamefont {Wang}, \citenamefont {Blumberger}, \citenamefont
  {Bonn},\ and\ \citenamefont
  {Beljonne}}]{giannini_TransientlyDelocalizedStates_2023}%
  \BibitemOpen
  \bibfield  {author} {\bibinfo {author} {\bibfnamefont {S.}~\bibnamefont
  {Giannini}}, \bibinfo {author} {\bibfnamefont {L.}~\bibnamefont
  {Di~Virgilio}}, \bibinfo {author} {\bibfnamefont {M.}~\bibnamefont
  {Bardini}}, \bibinfo {author} {\bibfnamefont {J.}~\bibnamefont {Hausch}},
  \bibinfo {author} {\bibfnamefont {J.}~\bibnamefont {Geuchies}}, \bibinfo
  {author} {\bibfnamefont {W.}~\bibnamefont {Zheng}}, \bibinfo {author}
  {\bibfnamefont {M.}~\bibnamefont {Volpi}}, \bibinfo {author} {\bibfnamefont
  {J.}~\bibnamefont {Elsner}}, \bibinfo {author} {\bibfnamefont
  {K.}~\bibnamefont {Broch}}, \bibinfo {author} {\bibfnamefont {Y.~H.}\
  \bibnamefont {Geerts}}, \bibinfo {author} {\bibfnamefont {F.}~\bibnamefont
  {Schreiber}}, \bibinfo {author} {\bibfnamefont {G.}~\bibnamefont
  {Schweicher}}, \bibinfo {author} {\bibfnamefont {H.~I.}\ \bibnamefont
  {Wang}}, \bibinfo {author} {\bibfnamefont {J.}~\bibnamefont {Blumberger}},
  \bibinfo {author} {\bibfnamefont {M.}~\bibnamefont {Bonn}},\ and\ \bibinfo
  {author} {\bibfnamefont {D.}~\bibnamefont {Beljonne}},\ }\Eprint
  {https://arxiv.org/abs/2303.13163} {arxiv:2303.13163 [cond-mat]}  (\bibinfo
  {year} {2023})\BibitemShut {NoStop}%
\bibitem [{\citenamefont {Zhou}\ and\ \citenamefont
  {Bernardi}(2019)}]{zhou_PredictingChargeTransport_2019}%
  \BibitemOpen
  \bibfield  {author} {\bibinfo {author} {\bibfnamefont {J.-J.}\ \bibnamefont
  {Zhou}}\ and\ \bibinfo {author} {\bibfnamefont {M.}~\bibnamefont
  {Bernardi}},\ }\href {https://doi.org/10.1103/PhysRevResearch.1.033138}
  {\bibfield  {journal} {\bibinfo  {journal} {Phys. Rev. Research}\ }\textbf
  {\bibinfo {volume} {1}},\ \bibinfo {pages} {033138} (\bibinfo {year}
  {2019})}\BibitemShut {NoStop}%
\bibitem [{\citenamefont {Mattoni}\ \emph {et~al.}(2015)\citenamefont
  {Mattoni}, \citenamefont {Filippetti}, \citenamefont {Saba},\ and\
  \citenamefont {Delugas}}]{mattoni_MethylammoniumRotationalDynamics_2015}%
  \BibitemOpen
  \bibfield  {author} {\bibinfo {author} {\bibfnamefont {A.}~\bibnamefont
  {Mattoni}}, \bibinfo {author} {\bibfnamefont {A.}~\bibnamefont {Filippetti}},
  \bibinfo {author} {\bibfnamefont {M.~I.}\ \bibnamefont {Saba}},\ and\
  \bibinfo {author} {\bibfnamefont {P.}~\bibnamefont {Delugas}},\ }\href
  {https://doi.org/10.1021/acs.jpcc.5b04283} {\bibfield  {journal} {\bibinfo
  {journal} {J. Phys. Chem. C}\ }\textbf {\bibinfo {volume} {119}},\ \bibinfo
  {pages} {17421} (\bibinfo {year} {2015})}\BibitemShut {NoStop}%
\bibitem [{\citenamefont {Hata}\ \emph {et~al.}(2017)\citenamefont {Hata},
  \citenamefont {Giorgi}, \citenamefont {Yamashita}, \citenamefont {Caddeo},\
  and\ \citenamefont {Mattoni}}]{hata_DevelopmentClassicalInteratomic_2017}%
  \BibitemOpen
  \bibfield  {author} {\bibinfo {author} {\bibfnamefont {T.}~\bibnamefont
  {Hata}}, \bibinfo {author} {\bibfnamefont {G.}~\bibnamefont {Giorgi}},
  \bibinfo {author} {\bibfnamefont {K.}~\bibnamefont {Yamashita}}, \bibinfo
  {author} {\bibfnamefont {C.}~\bibnamefont {Caddeo}},\ and\ \bibinfo {author}
  {\bibfnamefont {A.}~\bibnamefont {Mattoni}},\ }\href
  {https://doi.org/10.1021/acs.jpcc.6b11298} {\bibfield  {journal} {\bibinfo
  {journal} {J. Phys. Chem. C}\ }\textbf {\bibinfo {volume} {121}},\ \bibinfo
  {pages} {3724} (\bibinfo {year} {2017})}\BibitemShut {NoStop}%
\bibitem [{\citenamefont
  {Plimpton}(1995)}]{plimpton_FastParallelAlgorithms_1995}%
  \BibitemOpen
  \bibfield  {author} {\bibinfo {author} {\bibfnamefont {S.}~\bibnamefont
  {Plimpton}},\ }\href {https://doi.org/10.1006/jcph.1995.1039} {\bibfield
  {journal} {\bibinfo  {journal} {J. Comput. Phys.}\ }\textbf {\bibinfo
  {volume} {117}},\ \bibinfo {pages} {1} (\bibinfo {year} {1995})}\BibitemShut
  {NoStop}%
\bibitem [{SM()}]{SM}%
  \BibitemOpen
  \href@noop {} {}\bibinfo {note} {See Supplemental Material for theoretical
  and computational details}\BibitemShut {NoStop}%
\bibitem [{\citenamefont {Poglitsch}\ and\ \citenamefont
  {Weber}(1987)}]{poglitsch_DynamicDisorderMethylammoniumtrihalogenoplumbates_1987}%
  \BibitemOpen
  \bibfield  {author} {\bibinfo {author} {\bibfnamefont {A.}~\bibnamefont
  {Poglitsch}}\ and\ \bibinfo {author} {\bibfnamefont {D.}~\bibnamefont
  {Weber}},\ }\href {https://doi.org/10.1063/1.453467} {\bibfield  {journal}
  {\bibinfo  {journal} {J. Chem. Phys.}\ }\textbf {\bibinfo {volume} {87}},\
  \bibinfo {pages} {6373} (\bibinfo {year} {1987})}\BibitemShut {NoStop}%
\bibitem [{\citenamefont {Sharma}\ \emph {et~al.}(2020)\citenamefont {Sharma},
  \citenamefont {Dai}, \citenamefont {Gao}, \citenamefont {Brenner},
  \citenamefont {Yadgarov}, \citenamefont {Zhang}, \citenamefont {Rakita},
  \citenamefont {Korobko}, \citenamefont {Rappe},\ and\ \citenamefont
  {Yaffe}}]{sharma_ElucidatingAtomisticOrigin_2020}%
  \BibitemOpen
  \bibfield  {author} {\bibinfo {author} {\bibfnamefont {R.}~\bibnamefont
  {Sharma}}, \bibinfo {author} {\bibfnamefont {Z.}~\bibnamefont {Dai}},
  \bibinfo {author} {\bibfnamefont {L.}~\bibnamefont {Gao}}, \bibinfo {author}
  {\bibfnamefont {T.~M.}\ \bibnamefont {Brenner}}, \bibinfo {author}
  {\bibfnamefont {L.}~\bibnamefont {Yadgarov}}, \bibinfo {author}
  {\bibfnamefont {J.}~\bibnamefont {Zhang}}, \bibinfo {author} {\bibfnamefont
  {Y.}~\bibnamefont {Rakita}}, \bibinfo {author} {\bibfnamefont
  {R.}~\bibnamefont {Korobko}}, \bibinfo {author} {\bibfnamefont {A.~M.}\
  \bibnamefont {Rappe}},\ and\ \bibinfo {author} {\bibfnamefont
  {O.}~\bibnamefont {Yaffe}},\ }\href
  {https://doi.org/10.1103/PhysRevMaterials.4.092401} {\bibfield  {journal}
  {\bibinfo  {journal} {Phys. Rev. Mat.}\ }\textbf {\bibinfo {volume} {4}},\
  \bibinfo {pages} {092401} (\bibinfo {year} {2020})}\BibitemShut {NoStop}%
\bibitem [{\citenamefont {Mostofi}\ \emph {et~al.}(2014)\citenamefont
  {Mostofi}, \citenamefont {Yates}, \citenamefont {Pizzi}, \citenamefont {Lee},
  \citenamefont {Souza}, \citenamefont {Vanderbilt},\ and\ \citenamefont
  {Marzari}}]{mostofi_UpdatedVersionWannier90_2014}%
  \BibitemOpen
  \bibfield  {author} {\bibinfo {author} {\bibfnamefont {A.~A.}\ \bibnamefont
  {Mostofi}}, \bibinfo {author} {\bibfnamefont {J.~R.}\ \bibnamefont {Yates}},
  \bibinfo {author} {\bibfnamefont {G.}~\bibnamefont {Pizzi}}, \bibinfo
  {author} {\bibfnamefont {Y.-S.}\ \bibnamefont {Lee}}, \bibinfo {author}
  {\bibfnamefont {I.}~\bibnamefont {Souza}}, \bibinfo {author} {\bibfnamefont
  {D.}~\bibnamefont {Vanderbilt}},\ and\ \bibinfo {author} {\bibfnamefont
  {N.}~\bibnamefont {Marzari}},\ }\href
  {https://doi.org/10.1016/j.cpc.2014.05.003} {\bibfield  {journal} {\bibinfo
  {journal} {Comput. Phys. Commun.}\ }\textbf {\bibinfo {volume} {185}},\
  \bibinfo {pages} {2309} (\bibinfo {year} {2014})}\BibitemShut {NoStop}%
\bibitem [{\citenamefont {Kresse}\ and\ \citenamefont
  {Furthm{\"u}ller}(1996)}]{kresse_EfficientIterativeSchemes_1996}%
  \BibitemOpen
  \bibfield  {author} {\bibinfo {author} {\bibfnamefont {G.}~\bibnamefont
  {Kresse}}\ and\ \bibinfo {author} {\bibfnamefont {J.}~\bibnamefont
  {Furthm{\"u}ller}},\ }\href {https://doi.org/10.1103/PhysRevB.54.11169}
  {\bibfield  {journal} {\bibinfo  {journal} {Phys. Rev. B}\ }\textbf {\bibinfo
  {volume} {54}},\ \bibinfo {pages} {11169} (\bibinfo {year}
  {1996})}\BibitemShut {NoStop}%
\bibitem [{\citenamefont {Giannozzi}\ \emph {et~al.}(2020)\citenamefont
  {Giannozzi}, \citenamefont {Baseggio}, \citenamefont {Bonf{\`a}},
  \citenamefont {Brunato}, \citenamefont {Car}, \citenamefont {Carnimeo},
  \citenamefont {Cavazzoni}, \citenamefont {{de Gironcoli}}, \citenamefont
  {Delugas}, \citenamefont {Ferrari~Ruffino}, \citenamefont {Ferretti},
  \citenamefont {Marzari}, \citenamefont {Timrov}, \citenamefont {Urru},\ and\
  \citenamefont {Baroni}}]{giannozzi_UantumESPRESSOExascale_2020}%
  \BibitemOpen
  \bibfield  {author} {\bibinfo {author} {\bibfnamefont {P.}~\bibnamefont
  {Giannozzi}}, \bibinfo {author} {\bibfnamefont {O.}~\bibnamefont {Baseggio}},
  \bibinfo {author} {\bibfnamefont {P.}~\bibnamefont {Bonf{\`a}}}, \bibinfo
  {author} {\bibfnamefont {D.}~\bibnamefont {Brunato}}, \bibinfo {author}
  {\bibfnamefont {R.}~\bibnamefont {Car}}, \bibinfo {author} {\bibfnamefont
  {I.}~\bibnamefont {Carnimeo}}, \bibinfo {author} {\bibfnamefont
  {C.}~\bibnamefont {Cavazzoni}}, \bibinfo {author} {\bibfnamefont
  {S.}~\bibnamefont {{de Gironcoli}}}, \bibinfo {author} {\bibfnamefont
  {P.}~\bibnamefont {Delugas}}, \bibinfo {author} {\bibfnamefont
  {F.}~\bibnamefont {Ferrari~Ruffino}}, \bibinfo {author} {\bibfnamefont
  {A.}~\bibnamefont {Ferretti}}, \bibinfo {author} {\bibfnamefont
  {N.}~\bibnamefont {Marzari}}, \bibinfo {author} {\bibfnamefont
  {I.}~\bibnamefont {Timrov}}, \bibinfo {author} {\bibfnamefont
  {A.}~\bibnamefont {Urru}},\ and\ \bibinfo {author} {\bibfnamefont
  {S.}~\bibnamefont {Baroni}},\ }\href {https://doi.org/10.1063/5.0005082}
  {\bibfield  {journal} {\bibinfo  {journal} {J. Chem. Phys.}\ }\textbf
  {\bibinfo {volume} {152}},\ \bibinfo {pages} {154105} (\bibinfo {year}
  {2020})}\BibitemShut {NoStop}%
\bibitem [{\citenamefont {Oga}\ \emph {et~al.}(2014)\citenamefont {Oga},
  \citenamefont {Saeki}, \citenamefont {Ogomi}, \citenamefont {Hayase},\ and\
  \citenamefont {Seki}}]{oga_ImprovedUnderstandingElectronic_2014}%
  \BibitemOpen
  \bibfield  {author} {\bibinfo {author} {\bibfnamefont {H.}~\bibnamefont
  {Oga}}, \bibinfo {author} {\bibfnamefont {A.}~\bibnamefont {Saeki}}, \bibinfo
  {author} {\bibfnamefont {Y.}~\bibnamefont {Ogomi}}, \bibinfo {author}
  {\bibfnamefont {S.}~\bibnamefont {Hayase}},\ and\ \bibinfo {author}
  {\bibfnamefont {S.}~\bibnamefont {Seki}},\ }\href
  {https://doi.org/10.1021/ja506936f} {\bibfield  {journal} {\bibinfo
  {journal} {J. Am. Chem. Soc.}\ }\textbf {\bibinfo {volume} {136}},\ \bibinfo
  {pages} {13818} (\bibinfo {year} {2014})}\BibitemShut {NoStop}%
\bibitem [{\citenamefont {Savenije}\ \emph {et~al.}(2014)\citenamefont
  {Savenije}, \citenamefont {Ponseca}, \citenamefont {Kunneman}, \citenamefont
  {Abdellah}, \citenamefont {Zheng}, \citenamefont {Tian}, \citenamefont {Zhu},
  \citenamefont {Canton}, \citenamefont {Scheblykin}, \citenamefont
  {Pullerits}, \citenamefont {Yartsev},\ and\ \citenamefont
  {Sundstr{\"o}m}}]{savenije_ThermallyActivatedExciton_2014}%
  \BibitemOpen
  \bibfield  {author} {\bibinfo {author} {\bibfnamefont {T.~J.}\ \bibnamefont
  {Savenije}}, \bibinfo {author} {\bibfnamefont {C.~S.}\ \bibnamefont
  {Ponseca}}, \bibinfo {author} {\bibfnamefont {L.}~\bibnamefont {Kunneman}},
  \bibinfo {author} {\bibfnamefont {M.}~\bibnamefont {Abdellah}}, \bibinfo
  {author} {\bibfnamefont {K.}~\bibnamefont {Zheng}}, \bibinfo {author}
  {\bibfnamefont {Y.}~\bibnamefont {Tian}}, \bibinfo {author} {\bibfnamefont
  {Q.}~\bibnamefont {Zhu}}, \bibinfo {author} {\bibfnamefont {S.~E.}\
  \bibnamefont {Canton}}, \bibinfo {author} {\bibfnamefont {I.~G.}\
  \bibnamefont {Scheblykin}}, \bibinfo {author} {\bibfnamefont
  {T.}~\bibnamefont {Pullerits}}, \bibinfo {author} {\bibfnamefont
  {A.}~\bibnamefont {Yartsev}},\ and\ \bibinfo {author} {\bibfnamefont
  {V.}~\bibnamefont {Sundstr{\"o}m}},\ }\href
  {https://doi.org/10.1021/jz500858a} {\bibfield  {journal} {\bibinfo
  {journal} {J. Phys. Chem. Lett.}\ }\textbf {\bibinfo {volume} {5}},\ \bibinfo
  {pages} {2189} (\bibinfo {year} {2014})}\BibitemShut {NoStop}%
\bibitem [{\citenamefont {Milot}\ \emph {et~al.}(2015)\citenamefont {Milot},
  \citenamefont {Eperon}, \citenamefont {Snaith}, \citenamefont {Johnston},\
  and\ \citenamefont {Herz}}]{milot_TemperatureDependentCharge_2015}%
  \BibitemOpen
  \bibfield  {author} {\bibinfo {author} {\bibfnamefont {R.~L.}\ \bibnamefont
  {Milot}}, \bibinfo {author} {\bibfnamefont {G.~E.}\ \bibnamefont {Eperon}},
  \bibinfo {author} {\bibfnamefont {H.~J.}\ \bibnamefont {Snaith}}, \bibinfo
  {author} {\bibfnamefont {M.~B.}\ \bibnamefont {Johnston}},\ and\ \bibinfo
  {author} {\bibfnamefont {L.~M.}\ \bibnamefont {Herz}},\ }\href
  {https://doi.org/10.1002/adfm.201502340} {\bibfield  {journal} {\bibinfo
  {journal} {Adv. Funct. Mater.}\ }\textbf {\bibinfo {volume} {25}},\ \bibinfo
  {pages} {6218} (\bibinfo {year} {2015})}\BibitemShut {NoStop}%
\bibitem [{\citenamefont {Yi}\ \emph {et~al.}(2016)\citenamefont {Yi},
  \citenamefont {Wu}, \citenamefont {Zhu},\ and\ \citenamefont
  {Podzorov}}]{yi_IntrinsicChargeTransport_2016a}%
  \BibitemOpen
  \bibfield  {author} {\bibinfo {author} {\bibfnamefont {H.~T.}\ \bibnamefont
  {Yi}}, \bibinfo {author} {\bibfnamefont {X.}~\bibnamefont {Wu}}, \bibinfo
  {author} {\bibfnamefont {X.}~\bibnamefont {Zhu}},\ and\ \bibinfo {author}
  {\bibfnamefont {V.}~\bibnamefont {Podzorov}},\ }\href
  {https://doi.org/10.1002/adma.201600011} {\bibfield  {journal} {\bibinfo
  {journal} {Adv. Mater.}\ }\textbf {\bibinfo {volume} {28}},\ \bibinfo {pages}
  {6509} (\bibinfo {year} {2016})}\BibitemShut {NoStop}%
\bibitem [{\citenamefont {Shrestha}\ \emph {et~al.}(2018)\citenamefont
  {Shrestha}, \citenamefont {Matt}, \citenamefont {Osvet}, \citenamefont
  {Niesner}, \citenamefont {Hock},\ and\ \citenamefont
  {Brabec}}]{shrestha_AssessingTemperatureDependence_2018a}%
  \BibitemOpen
  \bibfield  {author} {\bibinfo {author} {\bibfnamefont {S.}~\bibnamefont
  {Shrestha}}, \bibinfo {author} {\bibfnamefont {G.~J.}\ \bibnamefont {Matt}},
  \bibinfo {author} {\bibfnamefont {A.}~\bibnamefont {Osvet}}, \bibinfo
  {author} {\bibfnamefont {D.}~\bibnamefont {Niesner}}, \bibinfo {author}
  {\bibfnamefont {R.}~\bibnamefont {Hock}},\ and\ \bibinfo {author}
  {\bibfnamefont {C.~J.}\ \bibnamefont {Brabec}},\ }\href
  {https://doi.org/10.1021/acs.jpcc.8b00341} {\bibfield  {journal} {\bibinfo
  {journal} {J. Phys. Chem. C}\ }\textbf {\bibinfo {volume} {122}},\ \bibinfo
  {pages} {5935} (\bibinfo {year} {2018})}\BibitemShut {NoStop}%
\bibitem [{\citenamefont {Xia}\ \emph {et~al.}(2021)\citenamefont {Xia},
  \citenamefont {Peng}, \citenamefont {Ponc{\'e}}, \citenamefont {Patel},
  \citenamefont {Wright}, \citenamefont {Crothers}, \citenamefont
  {Uller~Rothmann}, \citenamefont {Borchert}, \citenamefont {Milot},
  \citenamefont {Kraus}, \citenamefont {Lin}, \citenamefont {Giustino},
  \citenamefont {Herz},\ and\ \citenamefont
  {Johnston}}]{xia_LimitsElectricalMobility_2021}%
  \BibitemOpen
  \bibfield  {author} {\bibinfo {author} {\bibfnamefont {C.~Q.}\ \bibnamefont
  {Xia}}, \bibinfo {author} {\bibfnamefont {J.}~\bibnamefont {Peng}}, \bibinfo
  {author} {\bibfnamefont {S.}~\bibnamefont {Ponc{\'e}}}, \bibinfo {author}
  {\bibfnamefont {J.~B.}\ \bibnamefont {Patel}}, \bibinfo {author}
  {\bibfnamefont {A.~D.}\ \bibnamefont {Wright}}, \bibinfo {author}
  {\bibfnamefont {T.~W.}\ \bibnamefont {Crothers}}, \bibinfo {author}
  {\bibfnamefont {M.}~\bibnamefont {Uller~Rothmann}}, \bibinfo {author}
  {\bibfnamefont {J.}~\bibnamefont {Borchert}}, \bibinfo {author}
  {\bibfnamefont {R.~L.}\ \bibnamefont {Milot}}, \bibinfo {author}
  {\bibfnamefont {H.}~\bibnamefont {Kraus}}, \bibinfo {author} {\bibfnamefont
  {Q.}~\bibnamefont {Lin}}, \bibinfo {author} {\bibfnamefont {F.}~\bibnamefont
  {Giustino}}, \bibinfo {author} {\bibfnamefont {L.~M.}\ \bibnamefont {Herz}},\
  and\ \bibinfo {author} {\bibfnamefont {M.~B.}\ \bibnamefont {Johnston}},\
  }\href {https://doi.org/10.1021/acs.jpclett.1c00619} {\bibfield  {journal}
  {\bibinfo  {journal} {J. Phys. Chem. Lett.}\ }\textbf {\bibinfo {volume}
  {12}},\ \bibinfo {pages} {3607} (\bibinfo {year} {2021})}\BibitemShut
  {NoStop}%
\bibitem [{\citenamefont {Bruevich}\ \emph {et~al.}(2022)\citenamefont
  {Bruevich}, \citenamefont {Kasaei}, \citenamefont {Rangan}, \citenamefont
  {Hijazi}, \citenamefont {Zhang}, \citenamefont {Emge}, \citenamefont
  {Andrei}, \citenamefont {Bartynski}, \citenamefont {Feldman},\ and\
  \citenamefont {Podzorov}}]{bruevich_IntrinsicTrapFree_2022}%
  \BibitemOpen
  \bibfield  {author} {\bibinfo {author} {\bibfnamefont {V.}~\bibnamefont
  {Bruevich}}, \bibinfo {author} {\bibfnamefont {L.}~\bibnamefont {Kasaei}},
  \bibinfo {author} {\bibfnamefont {S.}~\bibnamefont {Rangan}}, \bibinfo
  {author} {\bibfnamefont {H.}~\bibnamefont {Hijazi}}, \bibinfo {author}
  {\bibfnamefont {Z.}~\bibnamefont {Zhang}}, \bibinfo {author} {\bibfnamefont
  {T.}~\bibnamefont {Emge}}, \bibinfo {author} {\bibfnamefont {E.~Y.}\
  \bibnamefont {Andrei}}, \bibinfo {author} {\bibfnamefont {R.~A.}\
  \bibnamefont {Bartynski}}, \bibinfo {author} {\bibfnamefont {L.~C.}\
  \bibnamefont {Feldman}},\ and\ \bibinfo {author} {\bibfnamefont
  {V.}~\bibnamefont {Podzorov}},\ }\href
  {https://doi.org/10.1002/adma.202205055} {\bibfield  {journal} {\bibinfo
  {journal} {Adv. Mater.}\ }\textbf {\bibinfo {volume} {34}},\ \bibinfo {pages}
  {2205055} (\bibinfo {year} {2022})}\BibitemShut {NoStop}%
\bibitem [{\citenamefont {Chen}\ \emph {et~al.}(2022)\citenamefont {Chen},
  \citenamefont {Motti}, \citenamefont {Oliver}, \citenamefont {Wright},
  \citenamefont {Snaith}, \citenamefont {Johnston}, \citenamefont {Herz},\ and\
  \citenamefont {Filip}}]{chen_OptoelectronicPropertiesMixed_2022}%
  \BibitemOpen
  \bibfield  {author} {\bibinfo {author} {\bibfnamefont {Y.}~\bibnamefont
  {Chen}}, \bibinfo {author} {\bibfnamefont {S.~G.}\ \bibnamefont {Motti}},
  \bibinfo {author} {\bibfnamefont {R.~D.~J.}\ \bibnamefont {Oliver}}, \bibinfo
  {author} {\bibfnamefont {A.~D.}\ \bibnamefont {Wright}}, \bibinfo {author}
  {\bibfnamefont {H.~J.}\ \bibnamefont {Snaith}}, \bibinfo {author}
  {\bibfnamefont {M.~B.}\ \bibnamefont {Johnston}}, \bibinfo {author}
  {\bibfnamefont {L.~M.}\ \bibnamefont {Herz}},\ and\ \bibinfo {author}
  {\bibfnamefont {M.~R.}\ \bibnamefont {Filip}},\ }\href
  {https://doi.org/10.1021/acs.jpclett.2c00938} {\bibfield  {journal} {\bibinfo
   {journal} {J. Phys. Chem. Lett.}\ }\textbf {\bibinfo {volume} {13}},\
  \bibinfo {pages} {4184} (\bibinfo {year} {2022})}\BibitemShut {NoStop}%
\bibitem [{\citenamefont {Chang}\ \emph {et~al.}(2004)\citenamefont {Chang},
  \citenamefont {Park},\ and\ \citenamefont
  {Matsuishi}}]{chang_FirstprinciplesStudyStructural_2004}%
  \BibitemOpen
  \bibfield  {author} {\bibinfo {author} {\bibfnamefont {Y.}~\bibnamefont
  {Chang}}, \bibinfo {author} {\bibfnamefont {C.~H.}\ \bibnamefont {Park}},\
  and\ \bibinfo {author} {\bibfnamefont {K.}~\bibnamefont {Matsuishi}},\
  }\href@noop {} {\bibfield  {journal} {\bibinfo  {journal} {J. Korean Phys.
  Soc.}\ }\textbf {\bibinfo {volume} {44}},\ \bibinfo {pages} {889} (\bibinfo
  {year} {2004})}\BibitemShut {NoStop}%
\bibitem [{\citenamefont {Feng}\ and\ \citenamefont
  {Xiao}(2014)}]{feng_CrystalStructuresOptical_2014}%
  \BibitemOpen
  \bibfield  {author} {\bibinfo {author} {\bibfnamefont {J.}~\bibnamefont
  {Feng}}\ and\ \bibinfo {author} {\bibfnamefont {B.}~\bibnamefont {Xiao}},\
  }\href {https://doi.org/10.1021/jz500480m} {\bibfield  {journal} {\bibinfo
  {journal} {J. Phys. Chem. Lett.}\ }\textbf {\bibinfo {volume} {5}},\ \bibinfo
  {pages} {1278} (\bibinfo {year} {2014})}\BibitemShut {NoStop}%
\bibitem [{\citenamefont {Chen}\ \emph {et~al.}(2015)\citenamefont {Chen},
  \citenamefont {De~Marco}, \citenamefont {Yang}, \citenamefont {Song},
  \citenamefont {Chen}, \citenamefont {Zhao}, \citenamefont {Hong},
  \citenamefont {Zhou},\ and\ \citenamefont
  {Yang}}]{chen_SpotlightOrganicInorganic_2015}%
  \BibitemOpen
  \bibfield  {author} {\bibinfo {author} {\bibfnamefont {Q.}~\bibnamefont
  {Chen}}, \bibinfo {author} {\bibfnamefont {N.}~\bibnamefont {De~Marco}},
  \bibinfo {author} {\bibfnamefont {Y.~M.}\ \bibnamefont {Yang}}, \bibinfo
  {author} {\bibfnamefont {T.-B.}\ \bibnamefont {Song}}, \bibinfo {author}
  {\bibfnamefont {C.-C.}\ \bibnamefont {Chen}}, \bibinfo {author}
  {\bibfnamefont {H.}~\bibnamefont {Zhao}}, \bibinfo {author} {\bibfnamefont
  {Z.}~\bibnamefont {Hong}}, \bibinfo {author} {\bibfnamefont {H.}~\bibnamefont
  {Zhou}},\ and\ \bibinfo {author} {\bibfnamefont {Y.}~\bibnamefont {Yang}},\
  }\href {https://doi.org/10.1016/j.nantod.2015.04.009} {\bibfield  {journal}
  {\bibinfo  {journal} {Nano Today}\ }\textbf {\bibinfo {volume} {10}},\
  \bibinfo {pages} {355} (\bibinfo {year} {2015})}\BibitemShut {NoStop}%
\bibitem [{\citenamefont {Galkowski}\ \emph {et~al.}(2016)\citenamefont
  {Galkowski}, \citenamefont {Mitioglu}, \citenamefont {Miyata}, \citenamefont
  {Plochocka}, \citenamefont {Portugall}, \citenamefont {Eperon}, \citenamefont
  {Wang}, \citenamefont {Stergiopoulos}, \citenamefont {Stranks}, \citenamefont
  {Snaith},\ and\ \citenamefont
  {Nicholas}}]{galkowski_DeterminationExcitonBinding_2016}%
  \BibitemOpen
  \bibfield  {author} {\bibinfo {author} {\bibfnamefont {K.}~\bibnamefont
  {Galkowski}}, \bibinfo {author} {\bibfnamefont {A.}~\bibnamefont {Mitioglu}},
  \bibinfo {author} {\bibfnamefont {A.}~\bibnamefont {Miyata}}, \bibinfo
  {author} {\bibfnamefont {P.}~\bibnamefont {Plochocka}}, \bibinfo {author}
  {\bibfnamefont {O.}~\bibnamefont {Portugall}}, \bibinfo {author}
  {\bibfnamefont {G.~E.}\ \bibnamefont {Eperon}}, \bibinfo {author}
  {\bibfnamefont {J.~T.-W.}\ \bibnamefont {Wang}}, \bibinfo {author}
  {\bibfnamefont {T.}~\bibnamefont {Stergiopoulos}}, \bibinfo {author}
  {\bibfnamefont {S.~D.}\ \bibnamefont {Stranks}}, \bibinfo {author}
  {\bibfnamefont {H.~J.}\ \bibnamefont {Snaith}},\ and\ \bibinfo {author}
  {\bibfnamefont {R.~J.}\ \bibnamefont {Nicholas}},\ }\href
  {https://doi.org/10.1039/C5EE03435C} {\bibfield  {journal} {\bibinfo
  {journal} {Energy Environ. Sci.}\ }\textbf {\bibinfo {volume} {9}},\ \bibinfo
  {pages} {962} (\bibinfo {year} {2016})}\BibitemShut {NoStop}%
\bibitem [{\citenamefont {Mosconi}\ \emph {et~al.}(2016)\citenamefont
  {Mosconi}, \citenamefont {Umari},\ and\ \citenamefont
  {De~Angelis}}]{mosconi_ElectronicOpticalProperties_2016}%
  \BibitemOpen
  \bibfield  {author} {\bibinfo {author} {\bibfnamefont {E.}~\bibnamefont
  {Mosconi}}, \bibinfo {author} {\bibfnamefont {P.}~\bibnamefont {Umari}},\
  and\ \bibinfo {author} {\bibfnamefont {F.}~\bibnamefont {De~Angelis}},\
  }\href {https://doi.org/10.1039/C6CP03969C} {\bibfield  {journal} {\bibinfo
  {journal} {Phys. Chem. Chem. Phys.}\ }\textbf {\bibinfo {volume} {18}},\
  \bibinfo {pages} {27158} (\bibinfo {year} {2016})}\BibitemShut {NoStop}%
\bibitem [{\citenamefont {Jong}\ \emph {et~al.}(2016)\citenamefont {Jong},
  \citenamefont {Yu}, \citenamefont {Ri}, \citenamefont {Kim},\ and\
  \citenamefont {Ri}}]{jong_InfluenceHalideComposition_2016}%
  \BibitemOpen
  \bibfield  {author} {\bibinfo {author} {\bibfnamefont {U.-G.}\ \bibnamefont
  {Jong}}, \bibinfo {author} {\bibfnamefont {C.-J.}\ \bibnamefont {Yu}},
  \bibinfo {author} {\bibfnamefont {J.-S.}\ \bibnamefont {Ri}}, \bibinfo
  {author} {\bibfnamefont {N.-H.}\ \bibnamefont {Kim}},\ and\ \bibinfo {author}
  {\bibfnamefont {G.-C.}\ \bibnamefont {Ri}},\ }\href
  {https://doi.org/10.1103/PhysRevB.94.125139} {\bibfield  {journal} {\bibinfo
  {journal} {Phys. Rev. B}\ }\textbf {\bibinfo {volume} {94}},\ \bibinfo
  {pages} {125139} (\bibinfo {year} {2016})}\BibitemShut {NoStop}%
\bibitem [{\citenamefont {Zu}\ \emph {et~al.}(2019)\citenamefont {Zu},
  \citenamefont {Amsalem}, \citenamefont {Egger}, \citenamefont {Wang},
  \citenamefont {Wolff}, \citenamefont {Fang}, \citenamefont {Loi},
  \citenamefont {Neher}, \citenamefont {Kronik}, \citenamefont {Duhm},\ and\
  \citenamefont {Koch}}]{zu_ConstructingElectronicStructure_2019a}%
  \BibitemOpen
  \bibfield  {author} {\bibinfo {author} {\bibfnamefont {F.}~\bibnamefont
  {Zu}}, \bibinfo {author} {\bibfnamefont {P.}~\bibnamefont {Amsalem}},
  \bibinfo {author} {\bibfnamefont {D.~A.}\ \bibnamefont {Egger}}, \bibinfo
  {author} {\bibfnamefont {R.}~\bibnamefont {Wang}}, \bibinfo {author}
  {\bibfnamefont {C.~M.}\ \bibnamefont {Wolff}}, \bibinfo {author}
  {\bibfnamefont {H.}~\bibnamefont {Fang}}, \bibinfo {author} {\bibfnamefont
  {M.~A.}\ \bibnamefont {Loi}}, \bibinfo {author} {\bibfnamefont
  {D.}~\bibnamefont {Neher}}, \bibinfo {author} {\bibfnamefont
  {L.}~\bibnamefont {Kronik}}, \bibinfo {author} {\bibfnamefont
  {S.}~\bibnamefont {Duhm}},\ and\ \bibinfo {author} {\bibfnamefont
  {N.}~\bibnamefont {Koch}},\ }\href
  {https://doi.org/10.1021/acs.jpclett.8b03728} {\bibfield  {journal} {\bibinfo
   {journal} {J. Phys. Chem. Lett.}\ }\textbf {\bibinfo {volume} {10}},\
  \bibinfo {pages} {601} (\bibinfo {year} {2019})}\BibitemShut {NoStop}%
\bibitem [{\citenamefont {Yamada}\ \emph {et~al.}(2021)\citenamefont {Yamada},
  \citenamefont {Mino}, \citenamefont {Kawahara}, \citenamefont {Oto},
  \citenamefont {Suzuura},\ and\ \citenamefont
  {Kanemitsu}}]{yamada_PolaronMassesCH_2021}%
  \BibitemOpen
  \bibfield  {author} {\bibinfo {author} {\bibfnamefont {Y.}~\bibnamefont
  {Yamada}}, \bibinfo {author} {\bibfnamefont {H.}~\bibnamefont {Mino}},
  \bibinfo {author} {\bibfnamefont {T.}~\bibnamefont {Kawahara}}, \bibinfo
  {author} {\bibfnamefont {K.}~\bibnamefont {Oto}}, \bibinfo {author}
  {\bibfnamefont {H.}~\bibnamefont {Suzuura}},\ and\ \bibinfo {author}
  {\bibfnamefont {Y.}~\bibnamefont {Kanemitsu}},\ }\href
  {https://doi.org/10.1103/PhysRevLett.126.237401} {\bibfield  {journal}
  {\bibinfo  {journal} {Phys. Rev. Lett.}\ }\textbf {\bibinfo {volume} {126}},\
  \bibinfo {pages} {237401} (\bibinfo {year} {2021})}\BibitemShut {NoStop}%
\bibitem [{\citenamefont {Traore}\ \emph {et~al.}(2022)\citenamefont {Traore},
  \citenamefont {Even}, \citenamefont {Pedesseau}, \citenamefont
  {K{\'e}p{\'e}nekian},\ and\ \citenamefont
  {Katan}}]{traore_BandGapEffective_2022}%
  \BibitemOpen
  \bibfield  {author} {\bibinfo {author} {\bibfnamefont {B.}~\bibnamefont
  {Traore}}, \bibinfo {author} {\bibfnamefont {J.}~\bibnamefont {Even}},
  \bibinfo {author} {\bibfnamefont {L.}~\bibnamefont {Pedesseau}}, \bibinfo
  {author} {\bibfnamefont {M.}~\bibnamefont {K{\'e}p{\'e}nekian}},\ and\
  \bibinfo {author} {\bibfnamefont {C.}~\bibnamefont {Katan}},\ }\href
  {https://doi.org/10.1103/PhysRevMaterials.6.014604} {\bibfield  {journal}
  {\bibinfo  {journal} {Phys. Rev. Materials}\ }\textbf {\bibinfo {volume}
  {6}},\ \bibinfo {pages} {014604} (\bibinfo {year} {2022})}\BibitemShut
  {NoStop}%
\bibitem [{\citenamefont {Filippetti}\ \emph {et~al.}(2016)\citenamefont
  {Filippetti}, \citenamefont {Mattoni}, \citenamefont {Caddeo}, \citenamefont
  {Saba},\ and\ \citenamefont
  {Delugas}}]{filippetti_LowElectronpolarOptical_2016a}%
  \BibitemOpen
  \bibfield  {author} {\bibinfo {author} {\bibfnamefont {A.}~\bibnamefont
  {Filippetti}}, \bibinfo {author} {\bibfnamefont {A.}~\bibnamefont {Mattoni}},
  \bibinfo {author} {\bibfnamefont {C.}~\bibnamefont {Caddeo}}, \bibinfo
  {author} {\bibfnamefont {M.~I.}\ \bibnamefont {Saba}},\ and\ \bibinfo
  {author} {\bibfnamefont {P.}~\bibnamefont {Delugas}},\ }\href
  {https://doi.org/10.1039/C6CP01402J} {\bibfield  {journal} {\bibinfo
  {journal} {Phys. Chem. Chem. Phys.}\ }\textbf {\bibinfo {volume} {18}},\
  \bibinfo {pages} {15352} (\bibinfo {year} {2016})}\BibitemShut {NoStop}%
\bibitem [{\citenamefont {Ponc{\'e}}\ \emph {et~al.}(2019)\citenamefont
  {Ponc{\'e}}, \citenamefont {Schlipf},\ and\ \citenamefont
  {Giustino}}]{ponce_OriginLowCarrier_2019a}%
  \BibitemOpen
  \bibfield  {author} {\bibinfo {author} {\bibfnamefont {S.}~\bibnamefont
  {Ponc{\'e}}}, \bibinfo {author} {\bibfnamefont {M.}~\bibnamefont {Schlipf}},\
  and\ \bibinfo {author} {\bibfnamefont {F.}~\bibnamefont {Giustino}},\ }\href
  {https://doi.org/10.1021/acsenergylett.8b02346} {\bibfield  {journal}
  {\bibinfo  {journal} {ACS Energy Lett.}\ }\textbf {\bibinfo {volume} {4}},\
  \bibinfo {pages} {456} (\bibinfo {year} {2019})}\BibitemShut {NoStop}%
\bibitem [{\citenamefont {Hellman}\ \emph {et~al.}(2013)\citenamefont
  {Hellman}, \citenamefont {Steneteg}, \citenamefont {Abrikosov},\ and\
  \citenamefont {Simak}}]{hellman_TemperatureDependentEffective_2013}%
  \BibitemOpen
  \bibfield  {author} {\bibinfo {author} {\bibfnamefont {O.}~\bibnamefont
  {Hellman}}, \bibinfo {author} {\bibfnamefont {P.}~\bibnamefont {Steneteg}},
  \bibinfo {author} {\bibfnamefont {I.~A.}\ \bibnamefont {Abrikosov}},\ and\
  \bibinfo {author} {\bibfnamefont {S.~I.}\ \bibnamefont {Simak}},\ }\href
  {https://doi.org/10.1103/PhysRevB.87.104111} {\bibfield  {journal} {\bibinfo
  {journal} {Phys. Rev. B}\ }\textbf {\bibinfo {volume} {87}},\ \bibinfo
  {pages} {104111} (\bibinfo {year} {2013})}\BibitemShut {NoStop}%
\bibitem [{\citenamefont {Klarbring}\ \emph {et~al.}(2020)\citenamefont
  {Klarbring}, \citenamefont {Hellman}, \citenamefont {Abrikosov},\ and\
  \citenamefont {Simak}}]{klarbring_AnharmonicityUltralowThermal_2020}%
  \BibitemOpen
  \bibfield  {author} {\bibinfo {author} {\bibfnamefont {J.}~\bibnamefont
  {Klarbring}}, \bibinfo {author} {\bibfnamefont {O.}~\bibnamefont {Hellman}},
  \bibinfo {author} {\bibfnamefont {I.~A.}\ \bibnamefont {Abrikosov}},\ and\
  \bibinfo {author} {\bibfnamefont {S.~I.}\ \bibnamefont {Simak}},\ }\href
  {https://doi.org/10.1103/PhysRevLett.125.045701} {\bibfield  {journal}
  {\bibinfo  {journal} {Phys. Rev. Lett.}\ }\textbf {\bibinfo {volume} {125}},\
  \bibinfo {pages} {045701} (\bibinfo {year} {2020})}\BibitemShut {NoStop}%
\end{thebibliography}%

\end{document}